\documentclass[twocolumn,showpacs,pra,floatfix]{revtex4}
\usepackage{graphicx,amsmath,amssymb,natbib,epsfig,multirow}
\newcommand\T{\rule{0pt}{2.6ex}}       
\newcommand\B{\rule[-1.2ex]{0pt}{0pt}} 

\graphicspath{{Figures/}} 

\begin{document}
\title{Collisions Between Ultracold Atoms and Cold Molecules \\in a Dual Electrostatic-Magnetic Trap}
\author{N. J. Fitch$^{1,3}$, L. P. Parazzoli$^{2,3}$, and H. J. Lewandowski$^3$}
\affiliation{1) Centre for Cold Matter, Blackett Laboratory, Imperial College London, Prince Consort Road, London SW7 2AZ, United Kingdom}
\affiliation{2) Sandia National Laboratories, Albuquerque, New Mexico 87185, USA}
\affiliation{3) JILA and Department of Physics, University of Colorado, Boulder, Colorado 80309, USA}
\date{\today}

\begin{abstract}
Measurements of interactions between cold molecules and ultracold atoms can allow for a detailed understanding of fundamental collision processes. These measurements can be done using various experimental geometries including where both species are in a beam, where one species is trapped, or when both species are trapped. Simultaneous trapping offers significantly longer interaction times and an associated increased sensitivity to rare collision events.  However, there are significant practical challenges associated with combining atom and molecule systems, which often have competing experimental requirements.
Here, we describe in detail an experimental system that allows for studies of cold collisions between ultracold atoms and cold molecules in a dual trap, where the atoms and molecules are trapped using static magnetic and electric fields, respectively. As a demonstration of the system's capabilities, we study cold collisions between ammonia ($^{14}$ND$_{3}$ and $^{15}$ND$_{3}$) molecules and rubidium ($^{87}$Rb and $^{85}$Rb) atoms.    
\end{abstract}

\pacs{37.10.Mn,37.10.20.+j,37.10.gh}

\maketitle       

\section{Introduction}

The study of interactions involving cold or ultracold neutral molecules is currently a research area of intense interest~\cite{Krems2008,Ye2009,Bohn2017}. Efforts in this area include studying elastic and inelastic scattering, as well as chemical reactions, all with the goal of understanding the fundamental quantum nature of the interactions. Examples of recent successes include the observation of light-assisted collisions of laser-cooled CaF molecules in optical tweezers~\cite{Anderegg2019}, control of chemical reactions between ultracold KRb molecules using stereodynamic effects~\cite{Miranda2011}, probing of quantum scattering resonances in Penning ionization~\cite{Henson2012,Jankunas2015,Klein2017} and reactions between excited metastable noble gas atoms and molecules~\cite{Gubbles2012b,Zou2019}, measurements of high-resolution differential cross-sections in crossed molecular beams of NO and H$_2$~\cite{Vogels2018}, and studies of dipolar collisions and reactions between polar molecules~\cite{Sawyer2011,Kirste2012,Gau2018,Wu2017,Segev2019,Hu2019}. Of particular relevance to the work presented here are experiments measuring collisions between trapped atoms and molecules, including N + NH~\cite{Hummon2011}, Li + O$_2$~\cite{Akerman2017}, and Rb + ND$_3$~\cite{Parazzoli2011}.

Various experimental architectures are used for investigating interactions, depending on the species under study, scientific goals, and temperature regimes.  For systems consisting of cold molecules, experiments typically employ merged or crossed molecular beams, a molecular beam impinging on a trapped sample, or a system where both species are trapped. The latter experiments take the form of either a co-trap, where the confining force for both species is created by the same field (i.e., electric, magnetic, or optical), or a dual-trap where the confining force for each species is created by a different field.  Each platform has its own advantages and limitations. Merged or crossed beams can allow for tuning of the collision energy with high resolution~\cite{Scharfenberg2011a,Chefdeville2012,Vogels2015,Poel2018,Zou2019}. In the case of merged beams, one can also study collisions at low temperatures~\cite{Henson2012,Shagam2013,Osterwalder2015,Amarasinghe2017}. However, beam-based experiments inherently have short interaction times, and thus less sensitivity for measuring rare collision events. There is also a limit to the lowest temperatures that can be probed based on the minimum beam forward speed for non-merged beam experiments. 

Trapping both species can enable studies with both low energies and long interaction times, allowing for high sensitivity to even very rare collision events.  Additionally, inelastic and elastic collisions can potentially be distinguished, as the former can result in an anit-trapped state that can be measured as trap loss, while the latter causes thermalizaton between the two species.  In the extreme limit of elastic collisions dominating the interactions, sympathetic cooling can be realized~cite{Monroe1993}.  This latter possibility makes trapped collisions particularly attractive as a pathway to creating ultracold molecules using the well-established techniques of ultracold atoms, but without the need for molecular laser cooling~\cite{Truppe2017,Anderegg2018,McCarron2018} or being constrained to bi-alkali species~\cite{Ni2008,Deiglmayr2008}.  Unfortunately, theoretical calculations using \textit{ab initio} potential energy surfaces are typically not accurate enough to be able to predict if a particular atom-molecular system will have a large enough ratio of elastic to inelastic collision cross-sections to facilitate sympathetic cooling.  Thus, direct experimental measurements are needed to constrain theoretical models.  Furthermore, carrying out experiments with different isotopologues is critical for constraining \textit{ab initio} preditions~\cite{Lavert2014}.  

Despite the advantages, there are also several challenges associated with using a trap to study collisions. First, since the inelastic scattering events happen over a long timescale and the resulting products are typically not trapped, it is not practical to directly measure the outgoing products or their quantum states. Thus, inelastic collisions are only indirectly inferred through measurements of trap loss.
Secondly, complex particle dynamics in a trap prohibit measurement of differential cross sections. Thirdly, the density distribution in the trap is not uniform in space or time, which adds temporal and spatial dependence to the collision rates.  Lastly, if the collision dynamics are affected by external fields, additional spatial dependence can come from the trapping fields themselves, which are necessarily spatially inhomogenous.  Such complications make the extraction of integrated cross sections difficult.  Nevertheless, the potential benefits of using a trap to study collisions make overcoming these challenges worthwhile.

The experimental approach and data analysis methods described in detail in this paper provide a platform for studying collisions between ultracold magnetically trapped atoms and cold electrostatically trapped polar molecules.  As a demonstration of the platform's capabilities, measurements of total inelastic and elastic cross sections between various isotopic combinations of ND$_{3}$ and Rb at collision energies of 100~mK are presented.  As mentioned above, inclusion of various isotopes in the collision experiments facilitates exploration of a overall scaling of the theoretical potential energy surface.  In this temperature regime, such a scaling could potentially alter the predicted collision cross sections by up to an order of magnitude~\cite{Frye2014}.  Here, we are able to explore collisions with multiple isotopologues of the molecules and isotopes of the atoms.  This paper is organized as follows.  In Section~\ref{Experiment}, the general approach to creating dual-trapped samples of ultracold Rb atoms and cold ND$_{3}$ molecules is described, including techniques used to characterize the two species and overlap the trapping potentials.  Section~\ref{Simulations} then describes trajectory simulations used to both provide insight into the complicated in-trap dynamics and extract collision cross sections.  Finally, Section~\ref{Results} presents collision results for the various isotope combinations.     

\section{Experiment}
\label{Experiment} 

\begin{figure}[!htb]
\centering
\includegraphics[width=\linewidth]{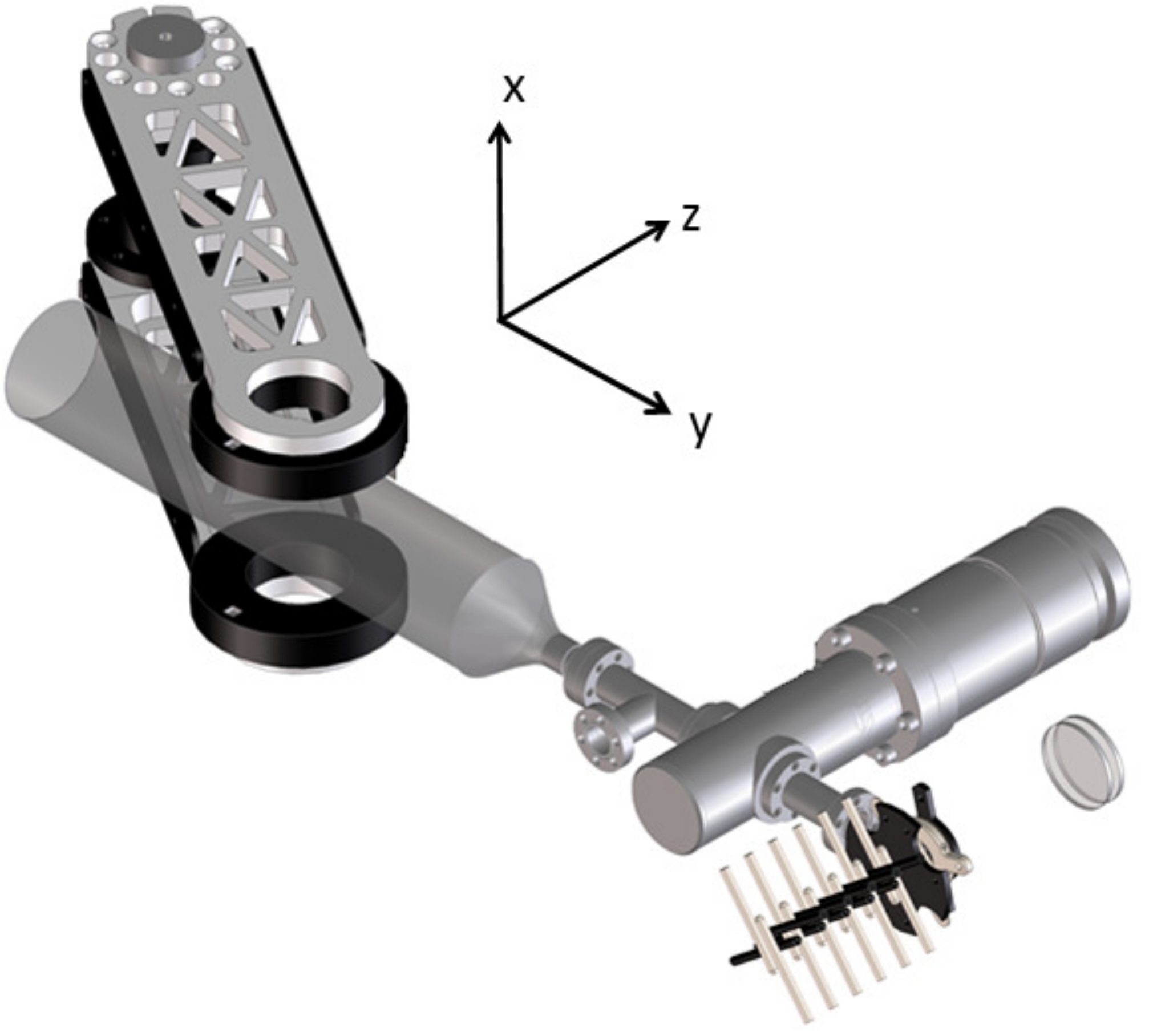}
\includegraphics[width=\linewidth]{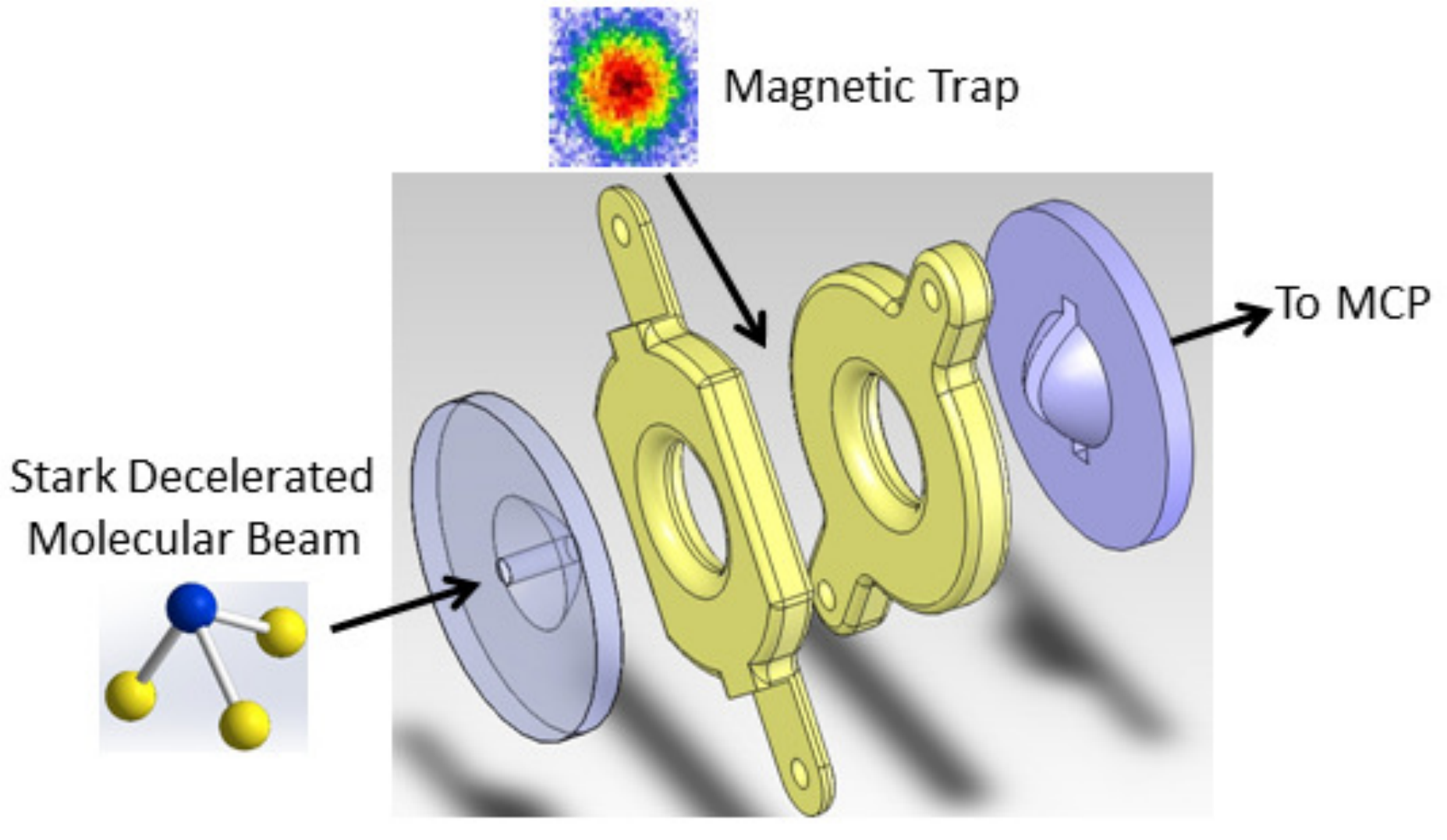}
\caption{Illustration of the experimental setup for the dual trap (not to scale).  (Top)  The electrostatic trap location, where the dual trap is formed, is located at the terminus of a Stark decelerator (alternating rod pairs) in the bottom right-hand corner.  The magnetic trap is formed between a translatable coil pair (dark disks mounted to long arms) external to the Rb vapor cell.  (Bottom) Closeup of the dual-trap region showing the four electrodes that form the electrostatic trap for the molecules. The distance between the electrodes has been increased for clarity.}
\label{Apparatus}
\end{figure}

A simplified illustration of the experimental apparatus is shown in Fig.~\ref{Apparatus}(Top).  The ultracold atom system consists of a cylindrical glass cell, anti-Helmholtz coils (black disks), and cooling lasers (not shown).  The cold molecule system consists of a pulsed valve (also not shown), Stark decelerator (alternating black/white rod pairs), and an electrostatic trap created by four electrodes mounted to the end of the decelerator.  An all-metal gate valve placed between the atom and molecule systems allows the two parts of the experiment to be vented or baked independently. The dual trap is created by first loading a magnetic trap of Rb from a magneto-optical trap and then translating the coil pair along the y-axis such that the trapped atoms are positioned at the center of the electrostatic trap electrodes, as shown using an exploded view of the latter in Fig.~\ref{Apparatus}(Bottom).  A gap of 5~mm between the two central electrodes allows the cloud of Rb atoms to enter the dual-trap region.  Once the atoms are in place, a pulsed molecular supersonic beam of ND$_3$ is created, decelerated, and trapped in the same physical location, forming the dual-trap environment.  Characterizations of both trapped species either before or after a controlled interaction time is accomplished using a combination of absorption imaging (for the atoms) and multi-photon ionization (for both species).  Ions are measured by biasing the electrostatic trap electrodes such that ions are extracted into a time-of-flight mass spectrometer with a two-stage microchannel plate (MCP) detector, shown as two gray discs next to the gate valve.    

\subsection{Creating Trapped Atoms and Molecules}
The general scheme for creating the magnetic trap of Rb atoms has been discussed in detail previously~\cite{Lewandowski2003}. The process begins by creating a magneto-optical trap (MOT) of either of the two most common isotopes of Rb, $^{87}$Rb or $^{85}$Rb.  The atoms are then optically pumped to a specific $|F,M_{F}\rangle$ hyperfine state, where $F$ and $M_{F}$ denote the total angular momentum and magnetic sublevel, respectively.  In order to optimize the loaded atom density, the atoms are trapped in their fully stretched states, $|F,M_{F}\rangle=|2,2\rangle$ for $^{87}$Rb, and $|F,M_{F}\rangle=|3,3\rangle$ for $^{85}$Rb.  Only these two states are included in this study, so the $|F,M_{F}\rangle$ labels are henceforth assumed for each isotope.  After optical pumping, the atoms are transferred into a magnetic trap with a field gradient of 375~G/cm in the strong dimension ($x$). The MOT and magnetic trapping fields are created by the same coils as shown in Fig.~\ref{Apparatus}(Top). Typical trapped atom populations consist of $\sim$10$^{9}$ atoms at a temperature of 600~$\mu$K and corresponding peak density of $\sim$10$^{10}$ atoms/cm$^{3}$.  Once the magnetic trap is loaded, the atoms can be transported through a differential pumping aperture to the dual-trap region by translating the track on which the coils are mounted. The aperture facilitates differential pumping and the required low pressure ( $\leq$1$\times$10$^{-9}$~torr) in the MOT chamber.

Production of the trapped cold molecules begins with creation of a pulsed supersonic beam sourced from a piezo-electric transducer (PZT) valve~\cite{Proch1989}.  The beam consists of either $^{14}$ND$_{3}$ or $^{15}$ND$_{3}$ seeded at a 1\% concentration in krypton, with a resulting mean forward speed of around 415~m/s.  The beam passes through a 2~mm diameter skimmer and is decelerated using a 149 stage Stark decelerator~\cite{Parazzoli2009}.  Here, spatially inhomogeneous, time-varying electric fields are used to decelerate a portion of the molecular beam.  The decelerated molecules are in the weak-field seeking $|j,k,m,\epsilon \rangle = |1,1,1,u \rangle$ quantum state, where $j$ is the total angular momentum, $k$ and $m$ are respectively the projections of $j$ onto the axis of symmetry and the laboratory field axis, and $u$ denotes that the molecules are in the weak-field seeking upper state of the inversion doublet~\cite{Townes}.  The slowed molecules, with a mean forward speed of approximately 26~m/s, exit the decelerator and are loaded into the electrostatic trap by removing the remainder of their forward kinetic energy using the trap electrodes as the final few stages of deceleration~\cite{Gilijamse2010,Gray2017}.  Once the molecules are brought to rest, the electrostatic trap is created by applying (typically) \{+8,-8,-8,+8\}~kV to the electrodes, as listed from the closest to farthest position in relation to the Stark decelerator.  With this configuration, the trap axis with the highest gradient corresponds to the molecular-beam axis, with a trap depth of approximately 600~mK and a potential energy gradient of 1.5~K/cm at the trap center. The radial dimensions (x,y) have a trap depth of approximately 140~mK, with a gradient of 0.4~K/cm at the trap center.  Approximately 10$^{4}$ molecules are trapped with characteristic energy widths of $\approx$100~mK and estimated peak densities of 10$^{6}$--10$^{7}$ molecules/cm$^{3}$.   

\subsection{Trapped Population Characterization}

A number of techniques are employed in order to understand the dynamics of both species in the dual trap, including ionization (for both atoms and molecules) and absorption (for atoms) detection methods.  The combination of approaches allows for accurate determination of the time dependence of the population and spatial distribution for both species in the dual trap.  The same measurements are also used to align the two independently trapped populations.

Properties of the trapped molecular samples are determined using a 2+1 resonantly enhanced multi-photon ionization (REMPI) scheme operated near 317~nm~\cite{Ashfold1987}.  The detection laser consists of a pulsed dye laser pumped by 532~nm light from a doubled Nd:YAG. The 10~mJ output pulse is focused by a 50~cm lens into the trap region with a beam waist of ($\approx$ 20~$\mu$m).  The resulting ions are extracted by rapidly switching the electrostatic trap electrodes into a \{1,1,0,0\}~kV configuration a few microseconds prior to triggering the ionization laser.  The resulting electric field accelerates the ions onto the MCP, effectively using the electrostatic trap as a time-of-flight mass spectrometer.  Because the REMPI laser beam waist in the detection region is significantly smaller than the characteristic trapped molecular cloud width (1~mm), the vertical ($x$) and horizontal ($z$) profiles of the molecular cloud can be measured by scanning the laser position\cite{Gray2017}.  A 2~mm wide and 1~cm tall slit in the last of the four trapping electrodes, as shown in Fig.~\ref{Apparatus}(Bottom), allows the ions to propagate through this electrode to the MCP.  Knowledge of the trapping potentials together with measurements of the spatial distribution of the molecules in the trap enables a rough estimate of the sample's temperature or energy distribution, though it is noted that the cloud is not in thermal equilibrium due to the lack of thermalizing elastic collisions in the low density sample.

\begin{figure}[!htb]
\centering
\includegraphics[width=\linewidth]{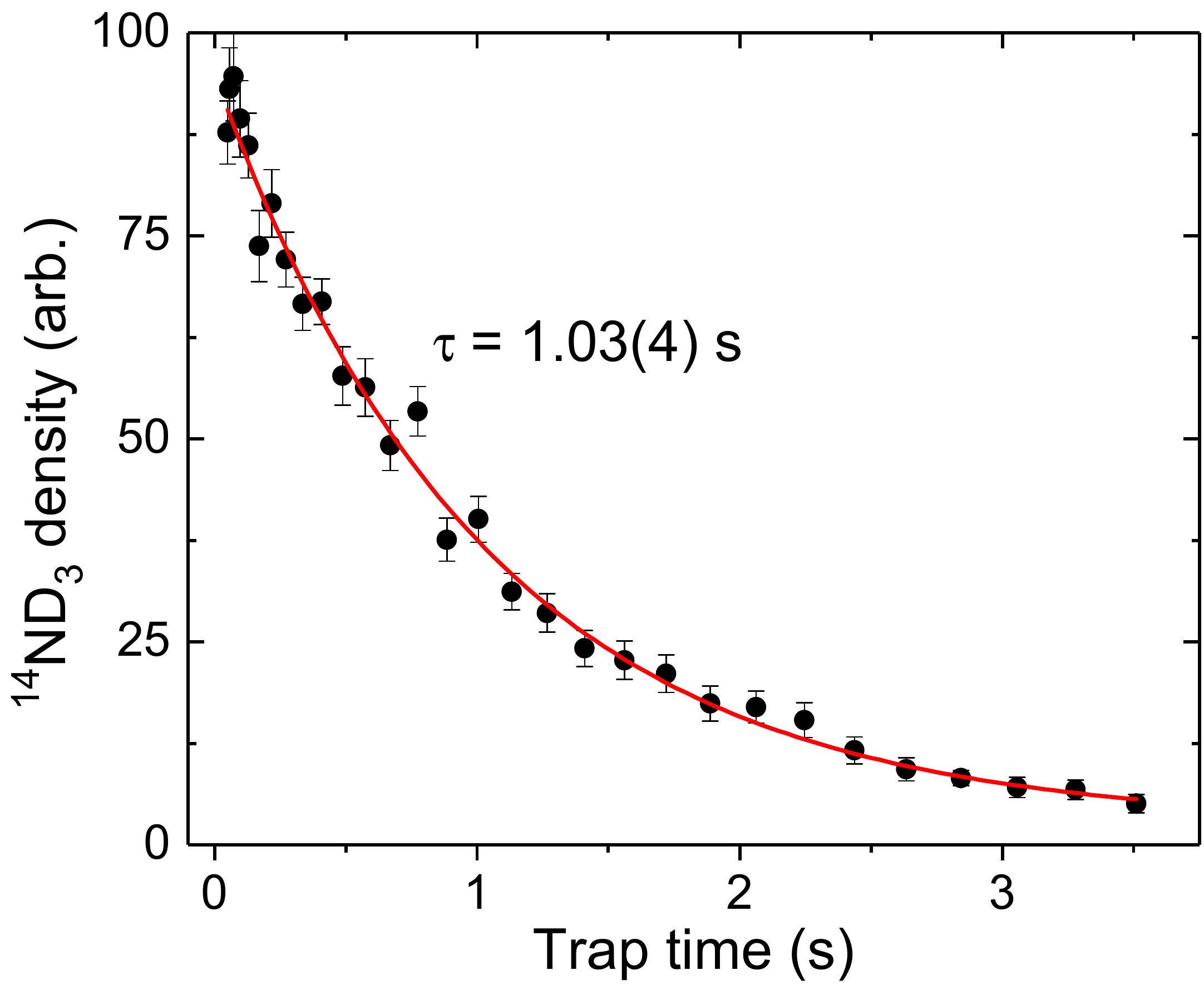}
\caption{$^{14}$ND$_{3}$ density at the trap center as a function of the trapping time (points).  Trapped populations of $^{15}$ND$_{3}$ display a similar density and trap lifetime.  The line represents a fit to a single exponential decay with a time constant $\tau$.}
\label{ND3_Decay_No_Rb}
\end{figure}
  
Using the REMPI laser, we can measure the ND$_{3}$ density at the trap center as a function of the trapping time as shown in Fig.~\ref{ND3_Decay_No_Rb} for $^{14}$ND$_{3}$.  Experimental measurements appear as black points, with statistical (1$\sigma$) error bars, and a single exponential fit appears as a solid line.  The fit indicates an in-trap 1/$e$ lifetime of approximately one second.  Data for $^{15}$ND$_{3}$, both in number and lifetime, are similar.  Loss of molecules from traps can typically be attributed to three main causes: optical pumping to untrapped states by blackbody radiation ~\cite{Hoekstra2007a}, nonadiabatic transitions at the trap center~\cite{Kirste2009}, and collisions with background gas.  The quality of the single exponential fit, as well as observations of a linear dependence of the lifetime on the chamber pressure, indicate that collisions with background gas are the dominant mechanism responsible for molecule loss in our experiment. 

Most properties of the trapped Rb cloud, including total number and temperature, are determined using absorption imaging.  Two independent physical locations are used for obtaining absorption images, one at the position of the dual trap and the other 20~cm away in the MOT cell.  Absorption images taken in the dual trap region are influenced by magnetic fields created by eddy currents in the steel vacuum chamber generated when the large magnetic trapping field is rapidly switched off.  These currents persist for a few milliseconds during which time the atoms are Zeeman-shifted out of resonance with the probe laser.  Quantitative measurements are therefore unreliable in this region for short expansion times.  As a result, absorption images taken in the dual trap environment are used only for rough diagnostics.  The secondary absorption image location in the MOT cell, which is constructed of glass, does not suffer from the same issues with eddy currents.  Therefore, the atoms are moved to this location for all measurements of the number and temperature.  No heating of the atom cloud during transport was detectable, making such a procedure possible.

\begin{figure}[!htb]
\centering
\includegraphics[width=\linewidth]{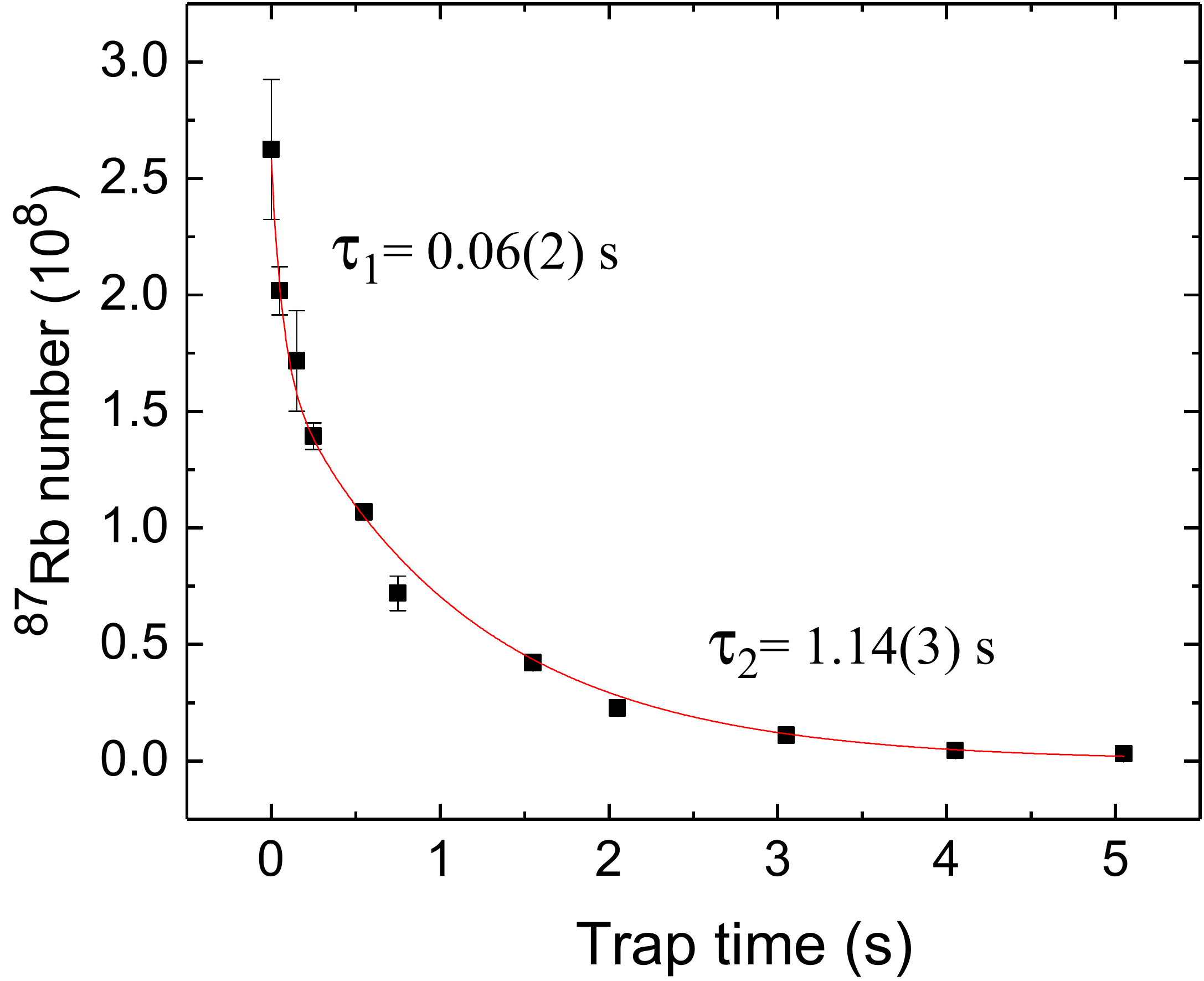}
\caption{Measured number of $^{87}$Rb atoms (black squares) after being exposed to the electrostatic trapping fields for varying amounts of time.  The observed decay fits well to a double exponential form (line).  The initial fast decay is due to loss of atoms by the reduction of the trapping potential from the electric field.  The slow decay is due to collisions with background gas.}
\label{Rb_Lifetime_In_Cotrap}
\end{figure} 

It is expected that the DC polarizablity Rb will cause a loss of Rb once the magnetic trap is overlapped with the electrostatic trap. Measurements of the total Rb number obtained via absorption images, as a function of time in the presence of the electrostatic trapping fields but without the presence of ND$_{3}$, appear in Fig.~\ref{Rb_Lifetime_In_Cotrap}.  The absorption measurements are all performed after the Rb is transported back to the MOT cell. We observe two distinct decay timescales, which we approximate by a double exponential decay of the form
\begin{equation}
N_{Rb}(t) = (N_{0}-N_{1})e^{-t\left(\frac{\tau_{1}+\tau_{2}}{\tau_{1} \tau_{2}}\right)} + N_{1}e^{-t/\tau_{2}},
\label{fitconstants}
\end{equation}
where $N_{Rb}(t)$ is the total number of Rb atoms present in the trapped cloud at time $t$, $N_0$ is the initial number of atoms, and $N_1$ is effectively the number of atoms that remain trapped in the presence of the electrostatic field. The initial fast decay ($\tau_{1}$) is due to a reduction of the trap depth due to the DC polarizablity of Rb once the large electrostatic trapping fields are applied. This decay mechanism turns off once the high-energy atoms have left the trap.  The slower decay is due to collisions with background gas.  If the electric fields are left off entirely, the population loss follows a single exponential decay with the longer of the two time constants ($\tau_{2}$).  Population behavior of both Rb isotopes is summarized in Table~\ref{Rb_Number_Decay_Table}.  We note that the experiment is optimized using $^{87}$Rb and the number of $^{85}$Rb atoms that we could reliably load into the magnetic trap was only half that of $^{87}$Rb.  This accounts for the factor of two discrepancy in initial number loaded.

\begin{table}[!htb]
\centering
\caption
{Rb number decay parameters in the dual trap region with $\pm$8~kV applied to the trap electrodes, as fit to equation~\ref{fitconstants}.}
\begin{tabular}{c|ccccc}
\hline
 & $N_{0}$~(10$^8$) & $N_{1}$~(10$^8$)  & $\tau_{1}$~(s) & $\tau_{2}$~(s) 
\\ \hline
 $^{87}$Rb & 2.6(5) & 1.7(1) & 0.06(2) & 1.14(3) \T \B
\\ \hline
 $^{85}$Rb & 1.5(3) & 0.8(2) & 0.05(2) & 0.90(1) \T \B
\\
\hline
\end{tabular}
\label{Rb_Number_Decay_Table}
\end{table} 

 As previously discussed, resonant absorption imaging in the dual-trap region is affected at short expansion times by eddy currents in the steel vacuum chamber.  Therefore, to collect in-trap dynamical information, we use the same laser that is used for REMPI detection of ND$_3$ to \textit{non-resonantly} ionize the trapped atoms while in the dual trap region.  Being a non-resonant process, this detection method is not sensitive to fields created by eddy currents as is the (resonant) absorption imaging.  Since the ionization energy of Rb is 4.18~eV, and the energy of a single REMPI photon is only 3.9~eV, this ionization process must involve multiple photons.  Nevertheless, due to the high intensity of the ionization laser, a substantial number of Rb$^{+}$ ions can be created.  This technique also enables measurements of the vertical Rb density profile during the interaction time, just as is done for the molecules.  The trap profile in the other two dimensions, which cannot be accurately measured experimentally, is determined using trajectory simulations of Rb atoms in the combined fields, as discussed in detail in Section \ref{Simulations}.   

\subsection{Dual Trap Alignment}
In addition to characterizing the in-trap dynamics and behavior of both species, the measurement techniques discussed thus far also facilitate spatial alignment between the two trapped populations.  A combination of methods are used for this purpose.  As the electrostatic trap location is fixed, adjustments of the dual trap overlap are made by adjusting the position of the magnetic trapping coils, which are located external to the vacuum chamber. The process to align the traps requires adjustment in all three dimensions and is an iterative process, as the alignment in one dimension is not always orthogonal to the other dimensions. 

\begin{figure}[!htb]
\centering
\includegraphics[width=0.85\linewidth]{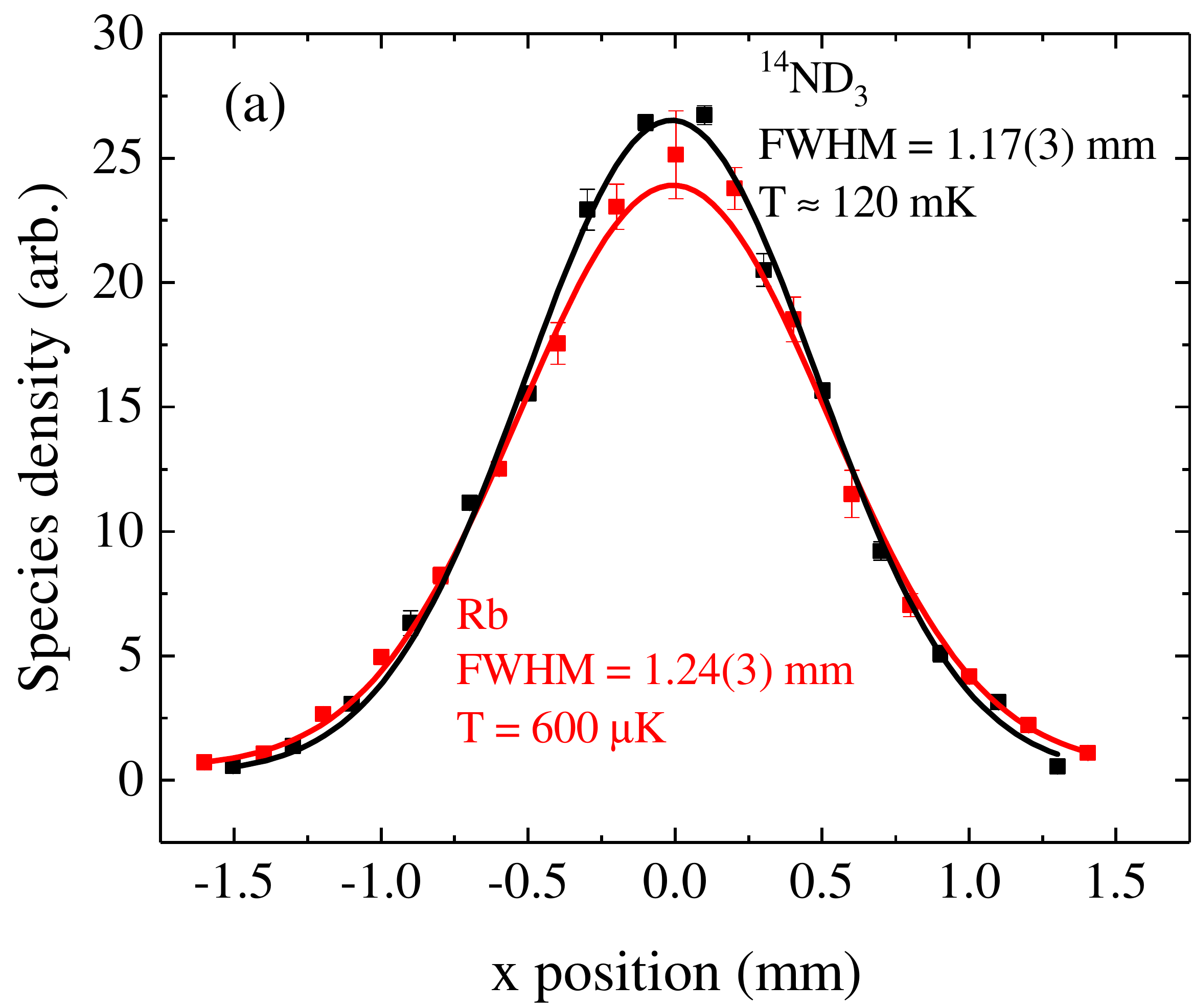}
\includegraphics[width=0.85\linewidth]{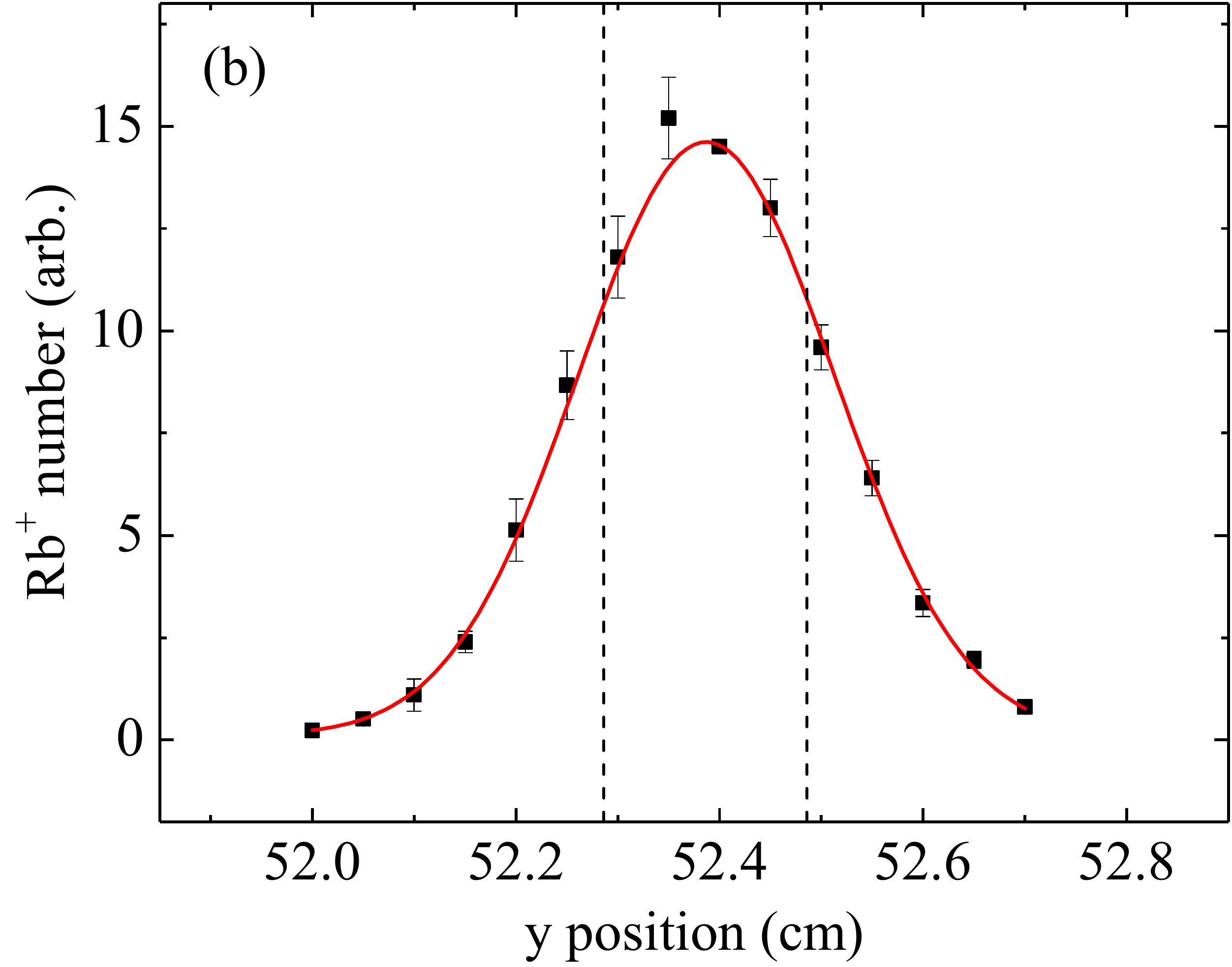}
\includegraphics[width=0.85\linewidth]{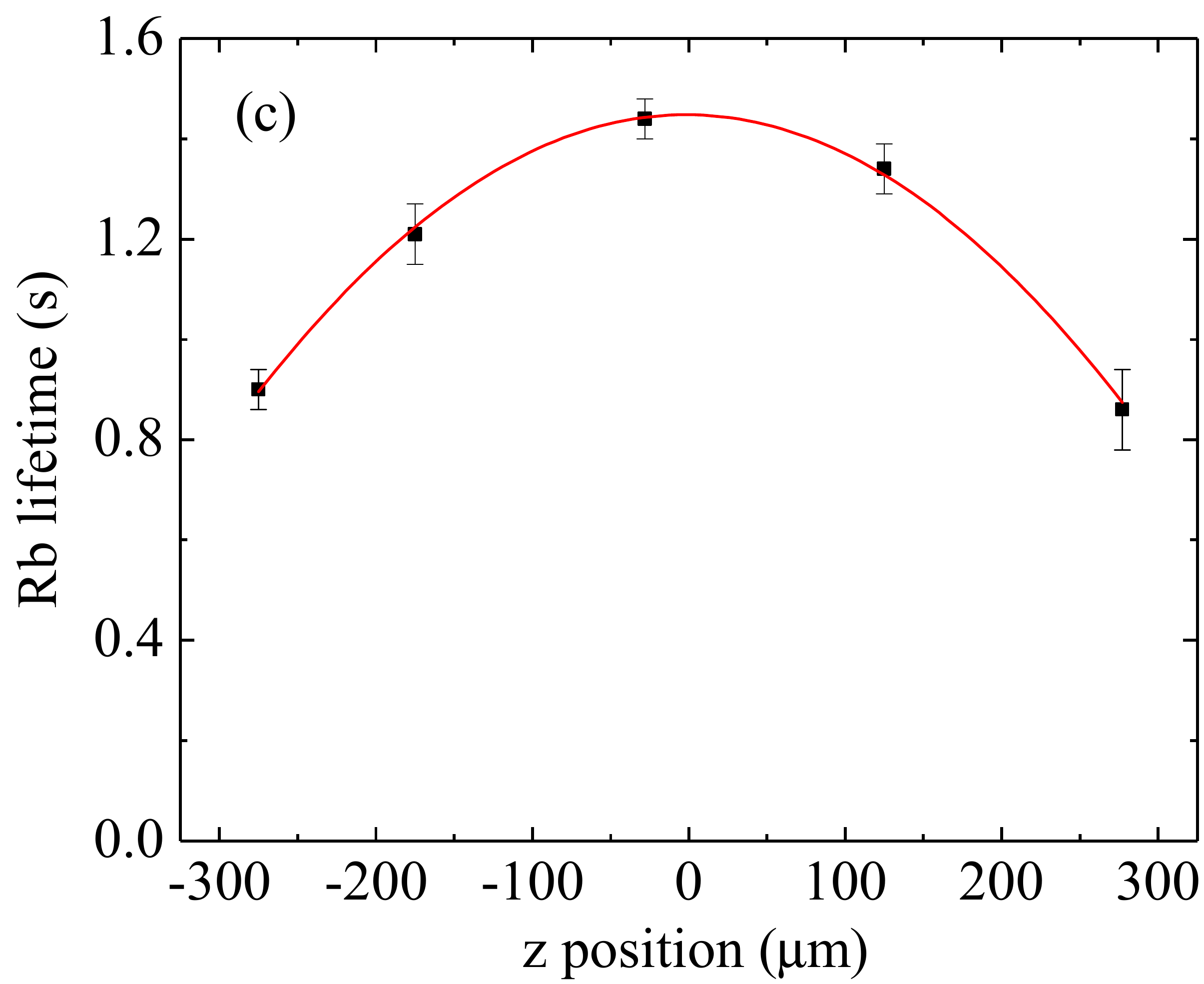}
\caption{Measurements used to align the two independent traps in all three spatial dimensions, with Gaussian fits (solid lines).  a) In the x-dimension, the profile of both Rb and ND$_{3}$ can be measured by scanning the ionization laser. The two data sets have been scaled to more easily compare the widths. b) In the y-dimension, the number of ionized Rb atoms are shown for various final track positions.  The vertical dashed lines correspond to the interpreted location of the slit extents in the back electrostatic-trap electrode.  c) In the z dimension, the Rb lifetime is maximized when the trapped atom cloud is aligned to the center of the electrostatic trapping fields.}
\label{Aligning}
\end{figure} 

The vertical ($x$)-dimension is aligned by performing scans of the ionization laser and overlapping the profiles of the molecule and atom samples, as illustrated in Fig.~\ref{Aligning}(a).  Adjustments to the height of the magnetic trap can be made by vertically translating the coil pair, or by introducing a slight imbalance in the two independently controlled coil currents.  Coincidentally, the atom and molecule traps have nearly identical widths in the vertical dimension.  This near-perfect spatial overlap is unique to this dimension, being the strong axis of the magnetic trap and a weak axis of the electrostatic trap.  The precision of the alignment in this dimension is limited by the uncertainties in the Gaussian fits, and is approximately 25~$\mu$m.  

The $y$ position of the magnetic trap corresponds to the dimension of the track on which the coils are mounted.  Moving the track to various final positions, and then non-resonantly ionizing Rb atoms in the trap, allows the $y$ position alignment to be optimized.  This approach is illustrated in Fig.~\ref{Aligning}(b).  The resulting Rb ion signal represents a convolution of the cloud shape and the extraction slit in the rearmost trap electrode.  Optimization of the number of Rb$^{+}$ ions arriving at the detector indicates optimal alignment.  The uncertainty of this alignment, approximately 50~$\mu$m, is again limited by the uncertainty in the Gaussian fit. The accuracy of the track positioning system itself is on the order of 10~$\mu$m, providing a lower limit on alignment precision in this dimension.  Alignment in this dimension can also be accomplished by taking absorption images of the Rb trap while the electric trapping fields are on.  Optimization of the number of atoms and symmetry of the image facilitate alignment.  However, this latter method was found to be less accurate, in part due to the previously discussed issues with eddy currents in the dual trap region.  

Alignment of the two traps along the longitudinal axis of the Stark decelerator ($z$) is the most challenging.  Here, we use the Rb lifetime in the dual trap to optimize the alignment, as shown in Fig.~\ref{Aligning}(c).  For these data, the Rb trap is pre-exposed to the electrostatic trapping fields for 0.5 seconds before lifetime measurements are taken and subsequently fit to a single exponential decay.  This approach reduces the sensitivity to the initially fast decay upon introduction to the electric fields ($\tau_{1}$ in equation~\ref{fitconstants}) and matches the conditions of the collision-data protocol (described below).  When optimally aligned, the Rb lifetime is maximized due to the Rb atoms experiencing the smallest possible electric field strengths.  Adjustments to the magnetic trap position along this axis are made by mechanically shimming the track on the optical table.  This position can be reproduced within approximately 50~$\mu$m.  As in the z dimension, absorption images in the dual-trap region can also be used to verify approximate alignment.  

\subsection{Collision Measurement Protocol}
The interactions between the two trapped species consist of both elastic and inelastic (internal energy changing) collisions.  Because the trapped atom number is $\sim$10$^{4}$ times that of the molecules, measuring changes in the Rb cloud due to the presence of the trapped ND$_{3}$ is impractical.  Hence, interactions are inferred by monitoring changes to the molecular cloud, typically measured at the trap center, with and without Rb atoms present.  Measuring the molecule density at the trap center most clearly reveals the impact the two collision types on the trapped sample.  Because the molecules are trapped in an excited weak-field-seeking state ($|1,1,1,u \rangle$), inelastic collisions can result in molecules being de-excited to untrapped states and lost.  At the low collision energies present in the these experiments, the only energetically allowed transitions are to the lower lying $|1,1,m,l \rangle$ or $|1,1,0,u \rangle$ states, neither of which are trapped.  At the same time as inelastic collisions lead to trap loss, elastic collisions tend to thermalize the two species.  Because the atoms are $\sim$200 times colder than the molecules, elastic collisions sympathetically cool the molecules and thus increase the molecule density at the trap center.  Competition between the two collision processes, in parallel with collisions with background gas, creates a time-dependent molecule density at the trap center.   

A typical experimental run proceeds as follows.  First, a Rb  MOT is created in the glass portion of the vacuum chamber.  A predetermined number of atoms, as measured by fluorescence imaging, are loaded into the trap. Loading a set number of atoms in the MOT reduces the sensitivity of the experiment to daily variations in MOT load time and final MOT population.  The atoms are then transferred into the purely magnetic trap and translated into the dual trap.  The shot-to-shot variation in the number of atoms transported to the interaction location is less than 10\%.  Once in the dual trap, the  Rb cloud is exposed to the electrostatic trapping fields for 0.1 seconds.  Although it reduces the number of Rb atoms available for collisions, this procedure reduces the sensitivity of the collision measurements to the initially fast dynamics of the Rb cloud upon exposure to the electric fields.  Next, the pulsed molecular beam is triggered and, within a few ms, a portion of the beam has been decelerated and brought to rest in the electrostatic trap. Collisions are then allowed to progress for a variable amount of time before the remaining molecule density at the trap center is measured via the 2+1 REMPI ionization scheme. 

Typically, a single shot of the experiment, including loading the MOT, transport of the magnetic trap, and collision interaction time, takes approximately 30 seconds.  The measurement is repeated 40 times for each interaction time.  To avoid systematic effects, data for various interaction times are taken in random order. Additionally, because background-gas collisions represent a significant amount of loss during the interaction time, we measure the difference between the ND$_{3}$ decay with and without Rb atoms present.  Measurements without Rb present are interleaved with each shot of the experiment.  For consistency between the two types of measurements, the experiment without Rb present is run in an identical manner as described above, except Rb is prevented from being loaded into the magnetic trap by blocking the repump laser during the MOT load time.  Finally, in order to maintain a constant average background pressure in the dual-trap region, the repetition rate of the experiment is the same for all interaction times.

\section{Extracting Cross Sections}
\label{Simulations}
As discussed above, the dual trap is a complex environment, where the population and spatial profile of both species are changing during the interaction time.  With so many dynamics occurring simultaneously, construction of an analytic model that accurately represents the ND$_{3}$ density decay is impractical.  Instead, we rely on trajectory simulations coupled with Monte Carlo techniques, which incorporate all observed trap dynamics, in order to model the trapped ND$_{3}$ decay and extract associated collision cross sections.  For each isotope combination, there are three simulations that are necessary to capture the full dynamics and to accurately model the experiment.  A preliminary simulation for each species is used to model the independent species behavior in the dual trap before a final simulation models the collisions.

\begin{figure}[!htb]
\centering
\includegraphics[width=0.9\linewidth]{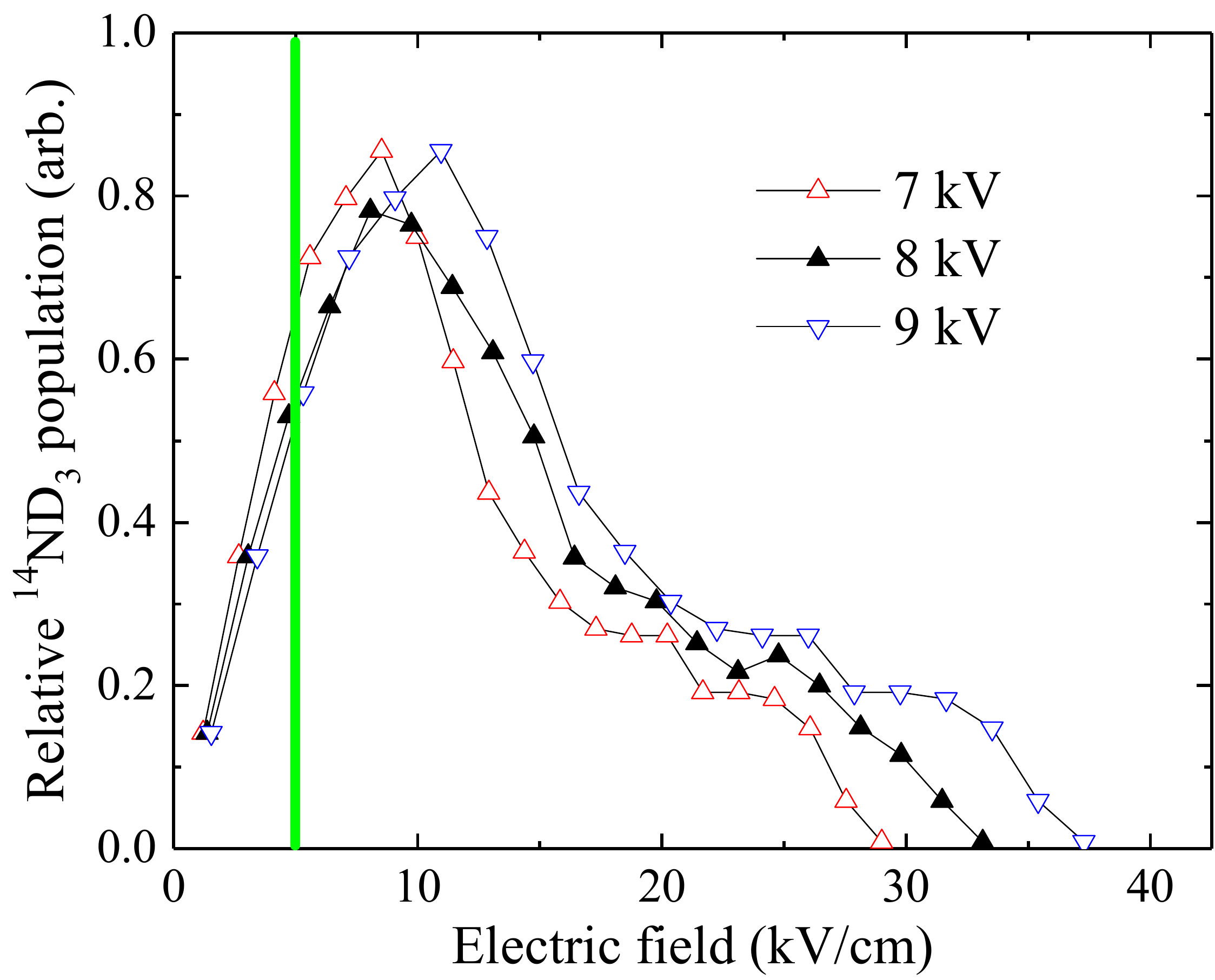}
\caption{Simulated average magnitude of the electric field sampled by molecules in the electrostatic trap for three different trap voltage amplitudes.  The vertical green line indicates the value for which the electric field enhancement of the inelastic cross section is expected to saturate.}
\label{Efield_Population_Distribution}
\end{figure}

The preliminary ND$_3$ trajectory simulation of the Stark decelerator is used to understand and optimize the deceleration and trap loading sequences and timings~\cite{Gilijamse2010}, as well as determine the starting phase-space distribution of the molecules in the trap. A model of the electric fields of the decelerator and trap is created using finite-element methods available from commercial software (COMSOL) and the molecules' trajectories are simulated through these fields using an extended Forest-Ruth like algorithm~\cite{Omelyan2002}.  Molecule loss due to background gas collisions is not directly simulated.  Instead, an overall exponential decay matching the experimentally measured value is applied. 
\begin{figure}[!htb]
\centering
\includegraphics[width=0.9\linewidth]{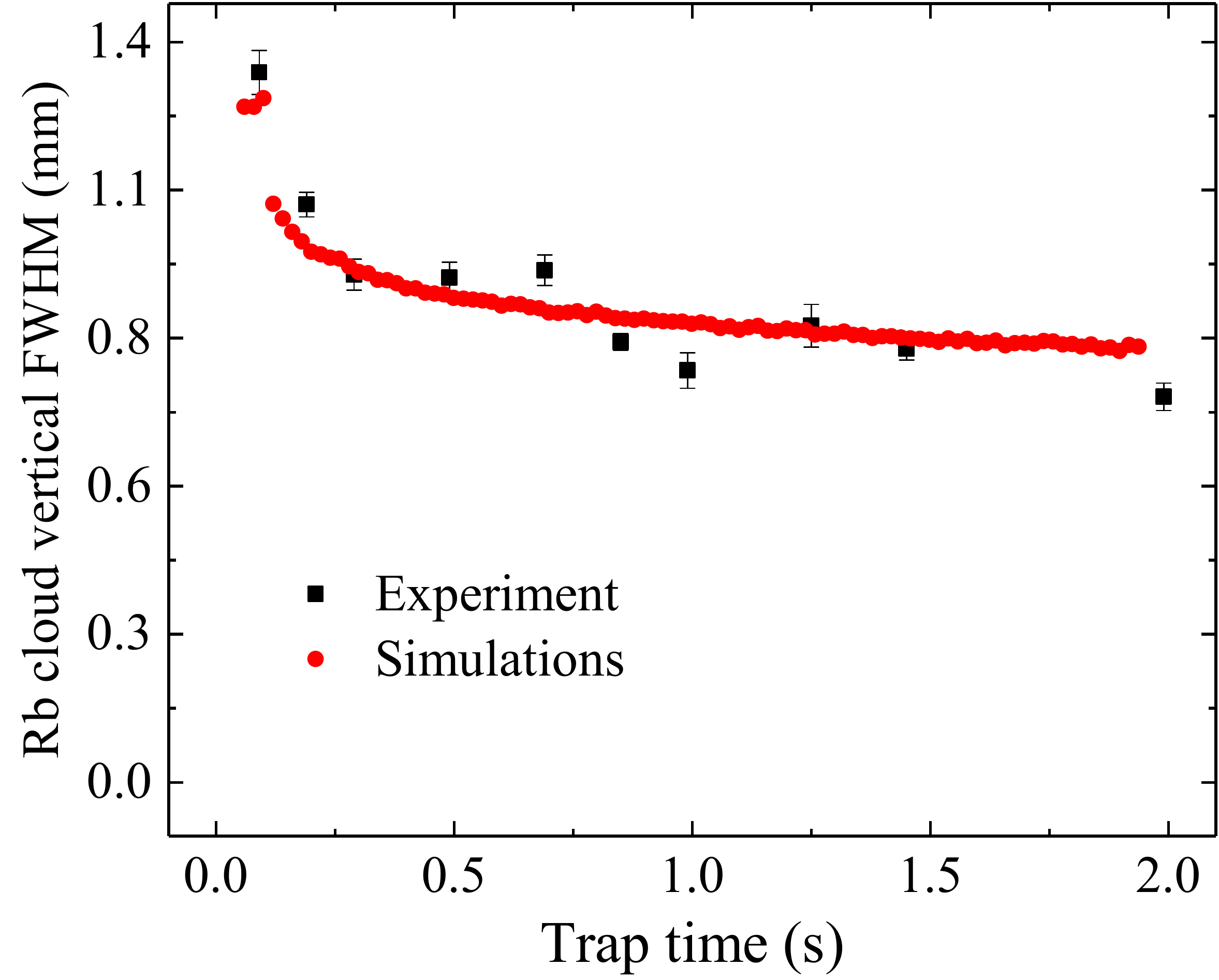}
\caption{Vertical (x) size of Rb cloud after the electrostatic trapping fields are turned on as determined by ionization detection (black points) and simulation (red points).}
\label{Rb_Vertical_Width_Data_Sims}
\end{figure} 

One of the crucial outputs of the preliminary ND$_{3}$ simulation is the average electric field experienced by the molecules, as shown in Fig.~\ref{Efield_Population_Distribution} for different trapping voltage amplitudes.  It is known that the collision cross section between Rb and ND$_{3}$ is electric field dependent, despite only one species being significantly polar~\cite{Zuchowski2008, Parazzoli2011}.  Specifically, the electric field enhances the inelastic collision rate while suppressing the elastic one. However, the electric-field effect saturates at about 5~kV/cm. In other words, the collision cross sections do not change above this field value.  The simulation results indicate that the vast majority of molecules experience average fields above 5~kV/cm.  Thus, in the final collision simulation, the cross sections can be treated as constants throughout the trap volume.

For the preliminary Rb simulation, the initial distribution is set to be that of a sample in thermal equilibrium in a quadrupole trap, with the number and temperature determined by absorption imaging.  The simulation begins with the electric trapping fields being turned on.  Elastic collisions between Rb atoms, which re-thermalize the cloud following the evaporation due to the lowering of the trapping potential from the interaction with the electric fields, are crucial in recreating the observed behavior.  Here, we assume purely s-wave collisions between Rb atoms using the field-free cross section appropriate for each isotope~\cite{Kempen2002}.  These simulations reproduce the experimentally observed double-exponential decay profile of the number of trapped atoms, as shown in Fig.~\ref{Rb_Lifetime_In_Cotrap}. Additionally, the simulations predict the behavior of the vertical Rb cloud width, which agrees reasonably well with the experimental measurements as shown in  Fig.~\ref{Rb_Vertical_Width_Data_Sims}.  These results give us confidence that the simulations accurately model the dynamics of the Rb cloud in the combined magnetic and electric fields.  
 
Next, we fit analytic models to the collective Rb cloud behavior (i.e., number and cloud widths as a function of time) and use these to model the Rb cloud as a mean field for the final collision simulations. This avoids the computationally intensive process of calculating individual Rb atom trajectories for the final collision simulation. This approximation is valid due to the large number of Rb atoms compared to ND$_{3}$, which means there is essentially no impact on the Rb cloud from atom-molecule collisions. Making this assumption, we model the  Rb density, $\rho_{Rb}$, as
\begin{equation}
\rho_{Rb} = \frac{N_{Rb}}{(2 \pi)^{3/2} \lambda_{x} \lambda_{y} \lambda_{z}} 
e^{
-\frac{1}{2} 
\left( \frac{x^{2}}{\lambda_{x}^{2}} + 
\frac{y^{2}}{\lambda_{y}^{2}} + 
\frac{z^{2}}{\lambda_{z}^{2}}
\right)
},
\end{equation}
where $\lambda_{i}$ is the time-dependent Gaussian trap width in the $i$ dimension ($i \in \left\{ x,y,z \right\}$).  The evolution of the cloud widths themselves are found to fit well to the form
\begin{equation}
\lambda_{i}(t)= 
\lambda_{1,i}e^{-t/\beta_{1,i}} + 
\lambda_{2,i}e^{-t/\beta_{2,i}} + 
\lambda_{3,i}.
\label{Rb_width_equation}
\end{equation} 
Fit parameter values for the width behavior of both isotopes of Rb appear in Table~\ref{Rb_Width_Decay_Table}.  The data presented is for an electrostatic trap voltage amplitude of 8~kV.  Separate simulations, resulting in altered fit parameters, are used for other trap voltages.  We note that the presence of the electrostatic trap electrodes breaks the y/z symmetry of the Rb cloud by restricting the cloud extents in the z dimension as the atoms are introduced into the dual trap via the 5~mm gap between the two central electrodes.    

\begin{table}[!htb]
\centering
\caption
{Parameters for Rb cloud width in the dual trap with 8~kV (amplitude) electrode potentials, as found by fitting simulation results to  Eqn.~\ref{Rb_width_equation}.  Length ($\lambda$) and time ($\beta$) units are mm and seconds, respectively.  Zero time is defined as when the electrostatic trapping fields are first applied.}
\begin{tabular}{cc|cccccc}
\hline %
& $i$ & $\lambda_{1}$ & $\beta_{1}$	& $\lambda_{2}$ & $\beta_{2}$ & $\lambda_{3}$ \\
\hline
\multirow{3}*{$^{87}$Rb} 
 	& $x$ & 0.18  & 0.07 & 0.08 & 0.75 & 0.35 \\
	& $y$ & 0.74  & 0.08 & 0.14 & 1.18 & 0.55 \\
 	& $z$ & 0.59  & 0.05 & 0.09 & 1.02 & 0.34 \\
\hline
\multirow{3}*{$^{85}$Rb} 
	& $x$ & 0.17 & 0.05 & 0.08 & 0.57 & 0.40 \\
	& $y$ & 0.93 & 0.05 & 0.17 & 0.65 & 0.61 \\
	& $z$ & 0.42 & 0.05 & 0.09 & 0.83 & 0.33 \\
\hline
\end{tabular}
\label{Rb_Width_Decay_Table}
\end{table} 

After the initial conditions for ND$_{3}$ molecules are well characterized and Rb is properly modeled as a mean field, the final collision simulation can be carried out. For the collision simulation, there are only two free parameters, the elastic and inelastic cross sections between Rb and ND$_3$.  During a simulation, the probability $P$ of an atom-molecule collision occurring for any individual ND$_{3}$ molecule is calculated at each time step $\delta t$ based on the local Rb density $\rho_{Rb}$, instantaneous molecular velocity $v$, and the assumed collision cross sections ($\sigma_{el}$ or $\sigma_{in}$), according to $P_{in/el} = \rho_{Rb}~\sigma_{in/el}~v~\delta t$.  To determine if a collision occurs for a particular time step, these probabilities are assigned to a unique subsection of a number line ranging between zero and one in proportion to their numeric value.  A random number is then chosen in this range.  If the random number is selected inside either collision range a collision of that type is assumed to have occurred.  If the random number falls outside either collision range, then no collision occurs.  $\delta t$ is chosen such that the local Rb density and molecular velocity are essentially constant over many time steps, and the collision probabilities are much less than unity.  Molecules that undergo an inelastic collision are simply removed from the simulation.  If a molecule undergoes an elastic collision, the recoil angle is chosen randomly from 4$\pi$ steradians.  ND$_{3}$--ND$_{3}$ collisions, which occur with frequencies on the order of 10$^{-5}$~Hz are neglected as they are insignificant compared to the  ND$_3$--Rb collisions,  which occur with frequencies on the order of 1 Hz.   

\begin{figure}[!htb]
\centering
\includegraphics[width=0.9\linewidth]{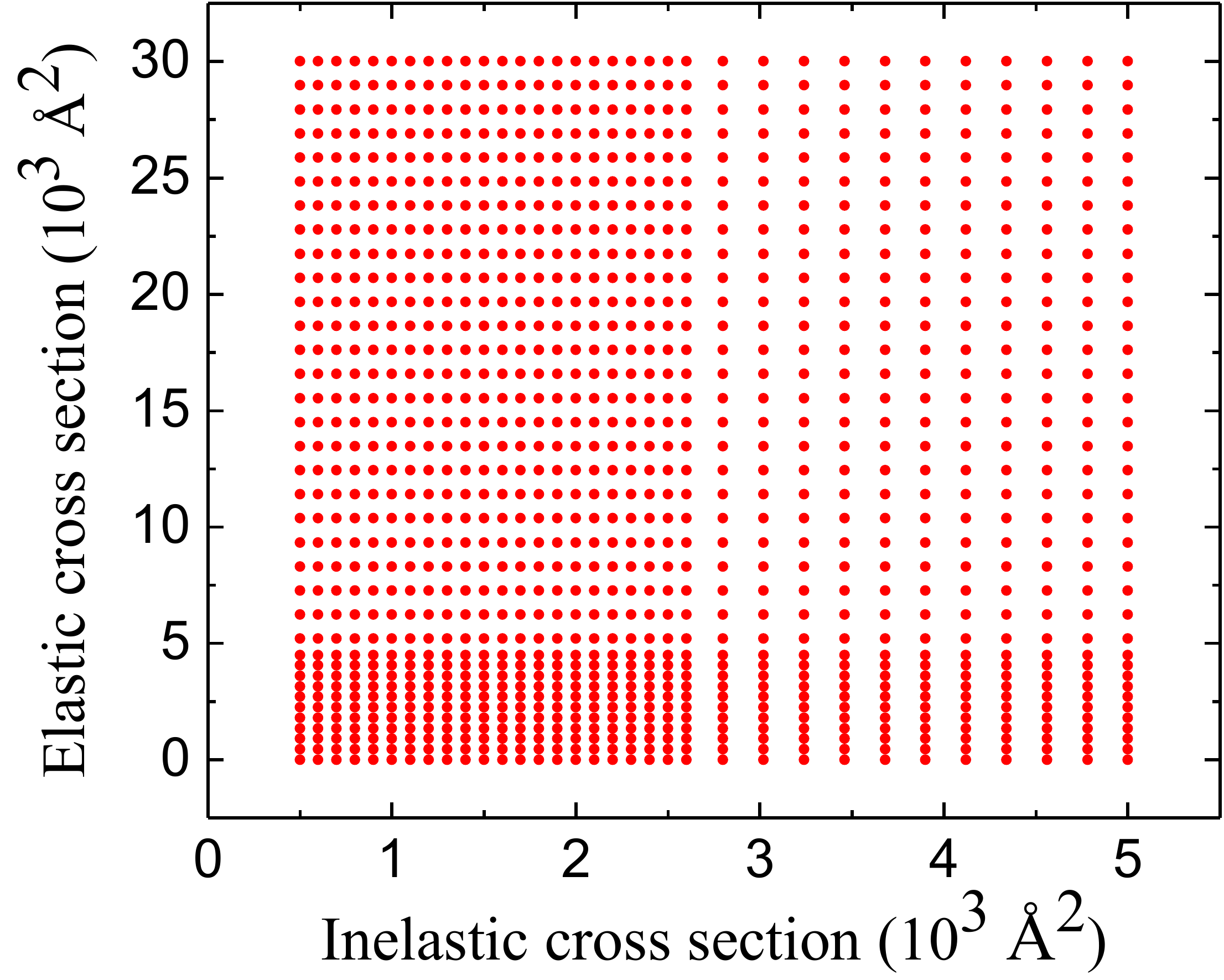}
\caption
{Parameter space of elastic and inelastic cross sections explored with the collision simulations.  A separate simulation is run for each point. The resulting simulated ND$_{3}$ peak density decay profile is quantitatively compared to the measured decay.}
\label{Simulation_Grid}
\end{figure}

The range of atom-molecule cross sections we explored is displayed in Fig.~\ref{Simulation_Grid}, where a separate collision simulation is run for each point in the grid. The reduced $\chi^{2}$ ~\cite{Cowen} is calculated for each simulation by comparing the simulated and experimentally measured decay of the peak ND$_3$ density (Fig.\ref{14ND3_87Rb_Collisions}(a)). Completing this process for all simulation points on the grid produces a contour plot indicating the most likely values of the cross sections (Fig. \ref{ND3_Rb_Contours}(a)).       

\section{Results and Discussion}
\label{Results}

\begin{figure}[!htb]
\centering
\includegraphics[width=0.85\linewidth]{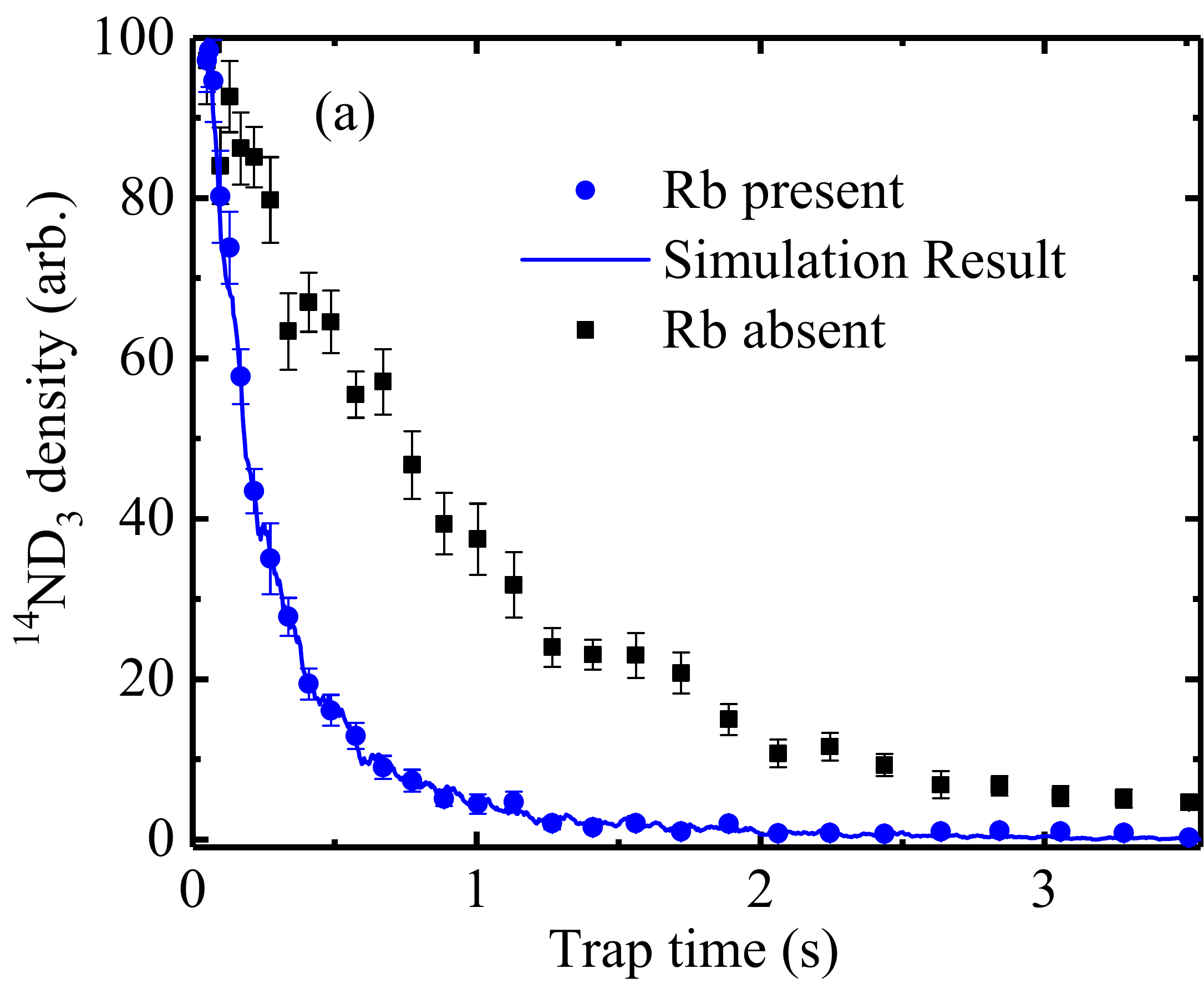}
\includegraphics[width=0.85\linewidth]{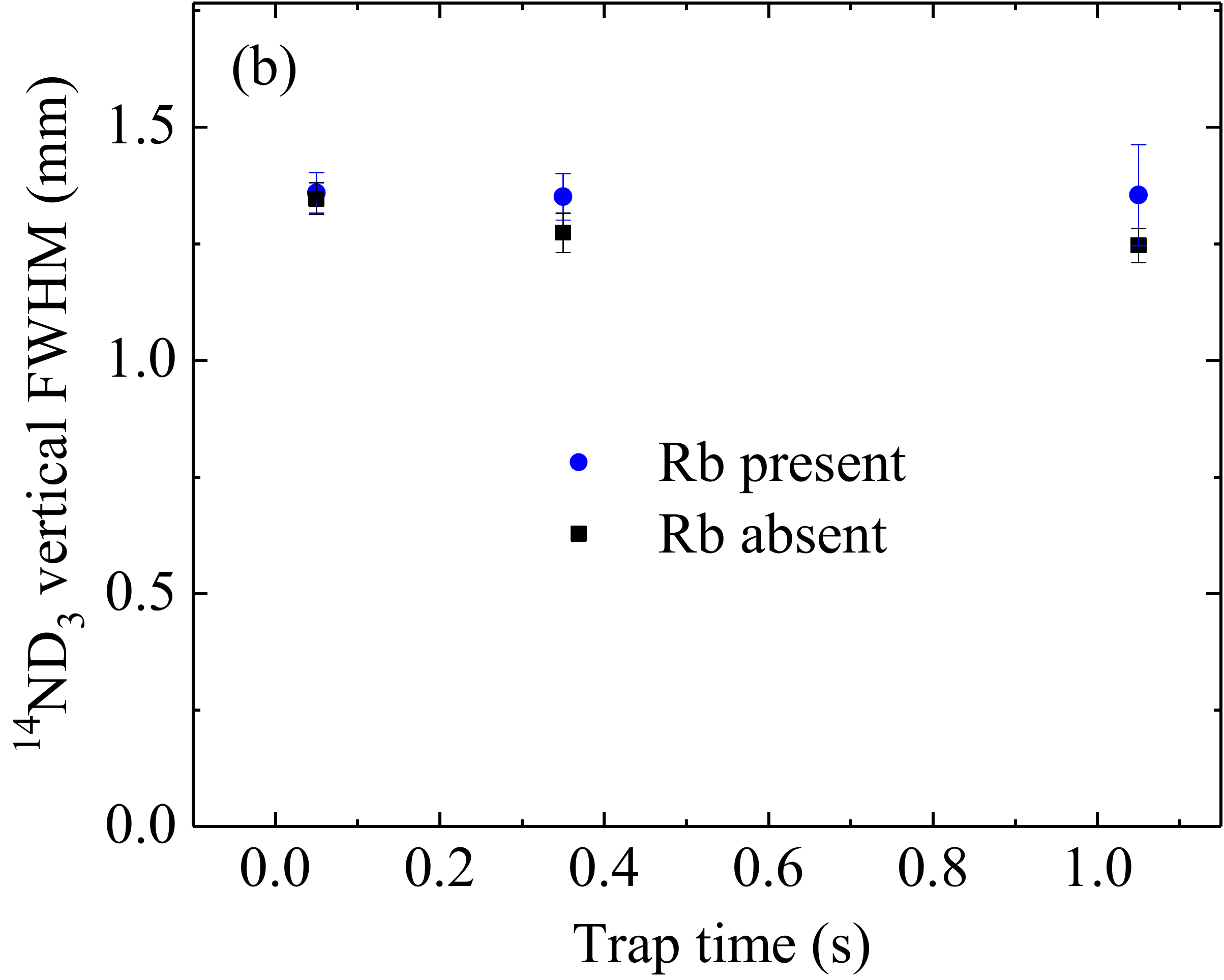}
\caption{Measurements of the trapped ND$_{3}$ peak density during the interaction time, both without (black squares) and with (blue points) Rb atoms present.  (a) $^{14}$ND$_{3}$ peak density along with best-fit simulation results (line).  (b) Measurements of the vertical (x) trapped molecular cloud width.}
\label{14ND3_87Rb_Collisions}
\end{figure}

Experimental data for measuring collisions between $^{14}$ND$_{3}$--$^{87}$Rb, along with best-fit simulation results corresponding to an elastic cross section of 2.5$\times$10$^{3}~$\AA$^{2}$ and an inelastic cross section of 2.0$\times$10$^{3}$~\AA$^{2}$, appear in Fig.~\ref{14ND3_87Rb_Collisions}. The ND$_3$ decay curves with (red squares) and without (black squares) trapped Rb present are shown. For the former, the starting Rb number and density, after an electric trap pre-exposure of 0.1 seconds, are 2.6$\times$10$^{8}$ and 8.1$\times$10$^{10}$ atoms/cm$^{3}$, respectively.  As shown, introducing Rb into the dual trap causes the molecule density at the trap center to decrease as a function of time, indicating that inelastic collisions cause molecule trap loss faster than elastic collisions can increase the molecule density. The simulations (solid curves) are able to reproduce the shape of the measured decay resulting from the competition between these processes.  In principle, experimental measurements of the trapped molecular cloud width might evolve during the interaction time. Such width measurements appear in Fig.~\ref{14ND3_87Rb_Collisions}(b) and show no measurable change during the collisions.  Thus, it is observed that measurements of the ND$_3$ density at the trap center is more sensitive to collisions than are cloud-density profile measurements.  

\begin{figure}[!htb]
\centering
\includegraphics[width=0.85\linewidth]{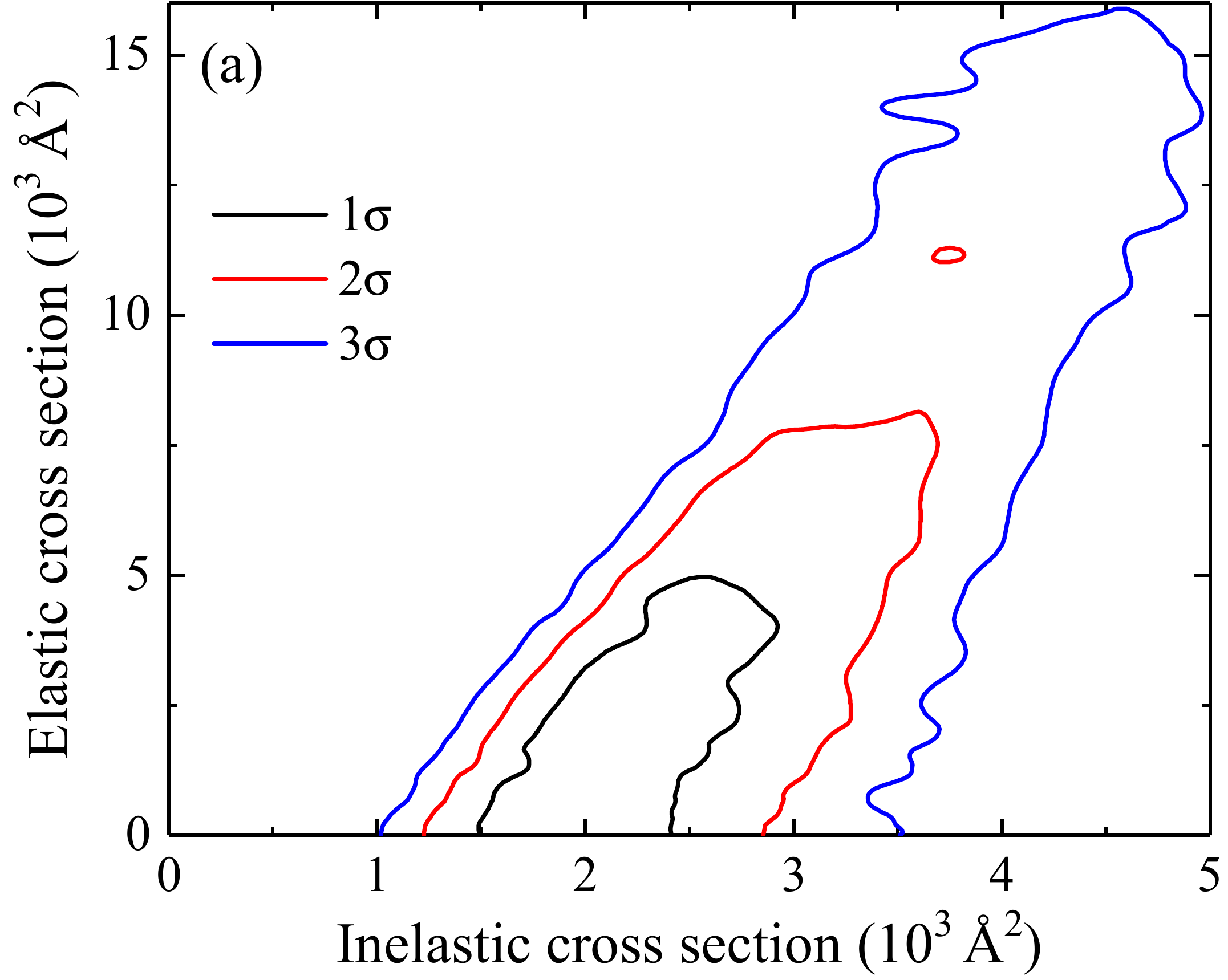}
\includegraphics[width=0.85\linewidth]{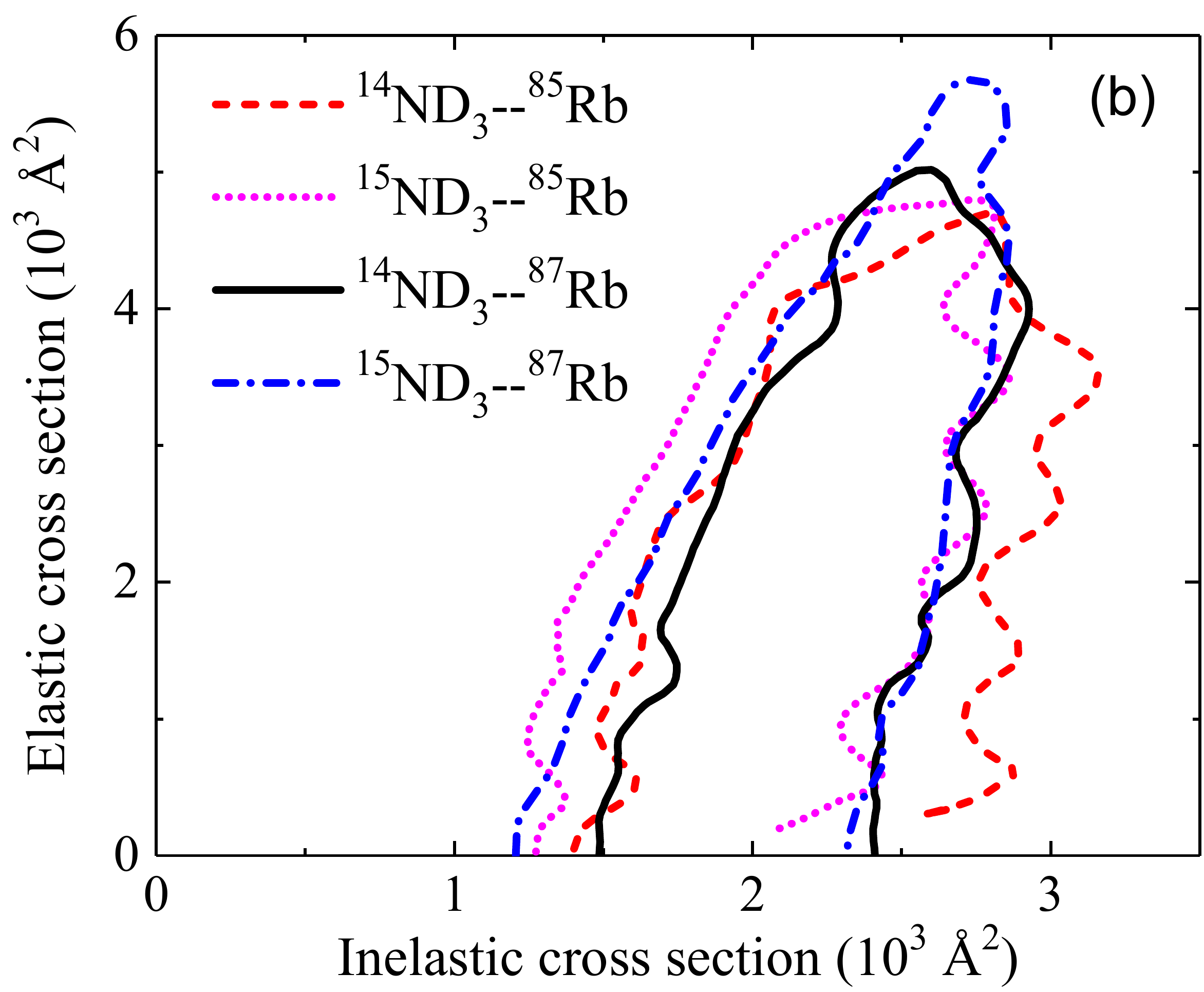}
\caption{(a) Confidence contours of the measured inelastic and elastic collision cross sections between $^{14}$ND$_{3}$ and $^{87}$Rb.  Due to the  ratio of $\sigma_{el}:\sigma_{in}$, we are capable of estimating a value for the inelastic cross section, but only an upper limit on the elastic cross section. (b) 1$\sigma$ confidence curves for various isotope combinations used in the experiment.}
\label{ND3_Rb_Contours}
\end{figure}

A contour plot of confidence intervals for the $^{14}$ND$_{3}$--$^{87}$Rb cross sections appears in Fig.~\ref{ND3_Rb_Contours}(a), including the 1$\sigma$ (68\%), 2$\sigma$ (95\%), and 3$\sigma$ (98\%) confidence bounds.  The contours are purely statistical and assume no uncertainty in measurement of the initial trap distributions.  From these data, an inelastic cross section of $\sigma_{in} \approx$ 2.0$\times$10$^{3}$~$\AA^{2}$ and $\sigma_{el} \leq$ 5.0$\times$10$^{3}$~$\AA^{2}$ can be assigned.  The elastic cross section is small enough that the contour plot does not close, indicating that the experiment is able to assign only an upper limit.  One general feature of the contour plots is a clear correlation between the elastic and inelastic cross sections.  This is due to the competition between the two collision processes on the density of the molecules at the trap center.  At larger elastic cross sections, more thermalization occurs, and the trap density at the center is increased.  A correspondingly larger inelastic cross section is required to reproduce the observed loss in molecule number at the trap center.

Similar contour plots for all four of the isotope combinations studied in this paper appear in Fig.~\ref{ND3_Rb_Contours}(b).  Only the 1$\sigma$ confidence bounds are shown for clarity.  For each of these curves, the behavior of both species is determined as above, and the entire set of simulations is rerun with the new inputs.  No significant differences are observed among the four isotope combinations.
 
\subsection{Robustness of cross-section assignments}

\begin{figure}[!htb]
\centering
\includegraphics[width=0.9\linewidth]{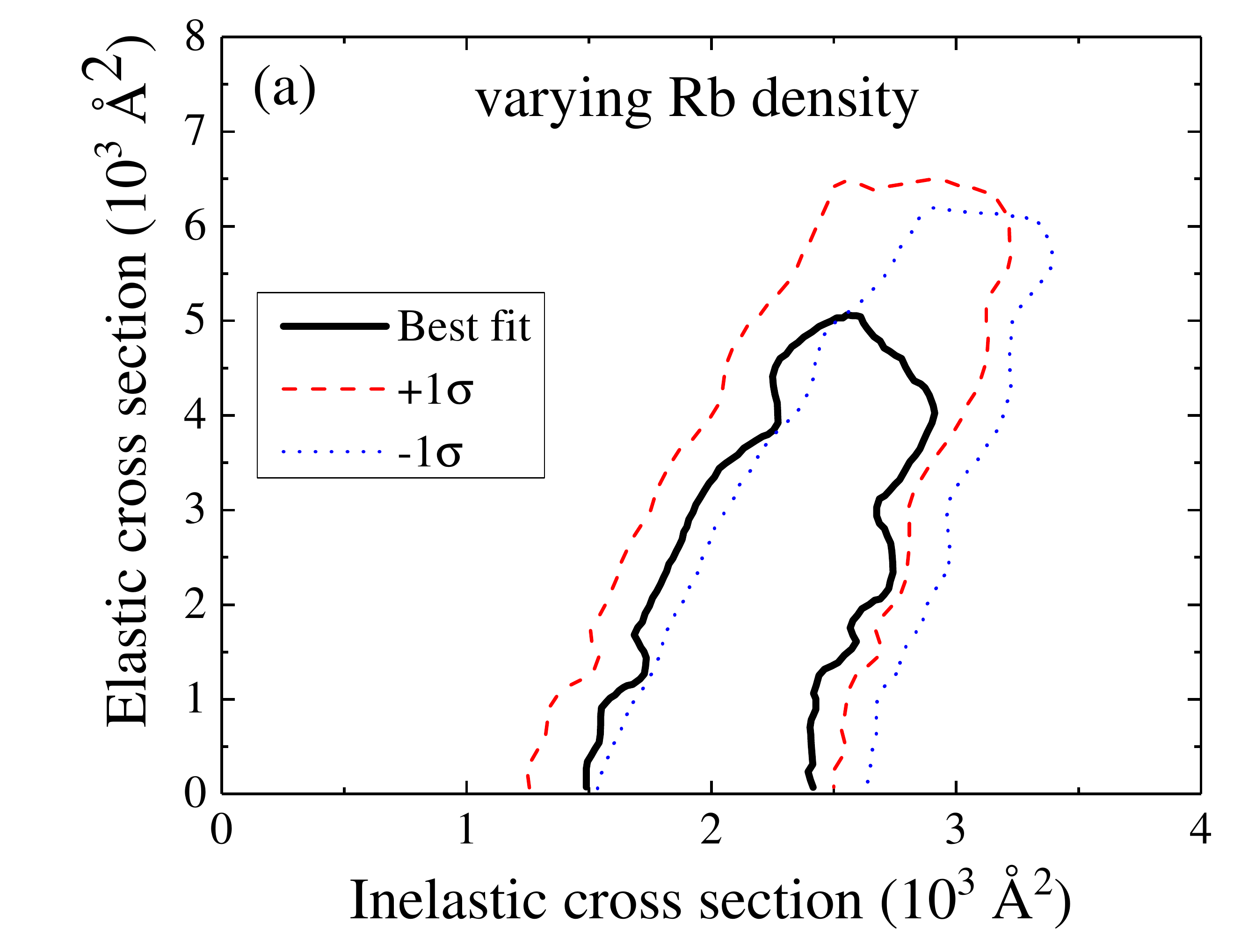}
\includegraphics[width=0.9\linewidth]{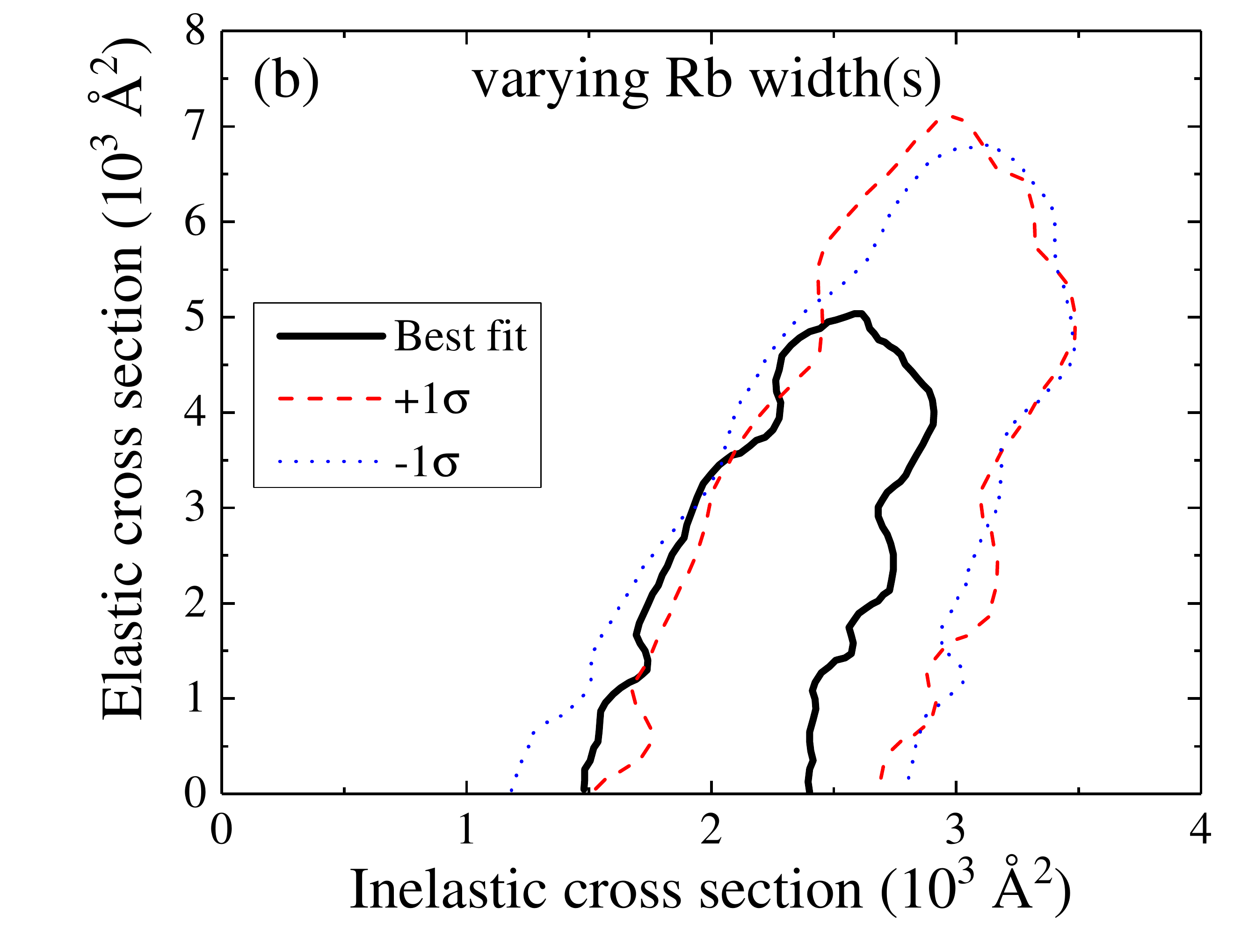}
\includegraphics[width=0.9\linewidth]{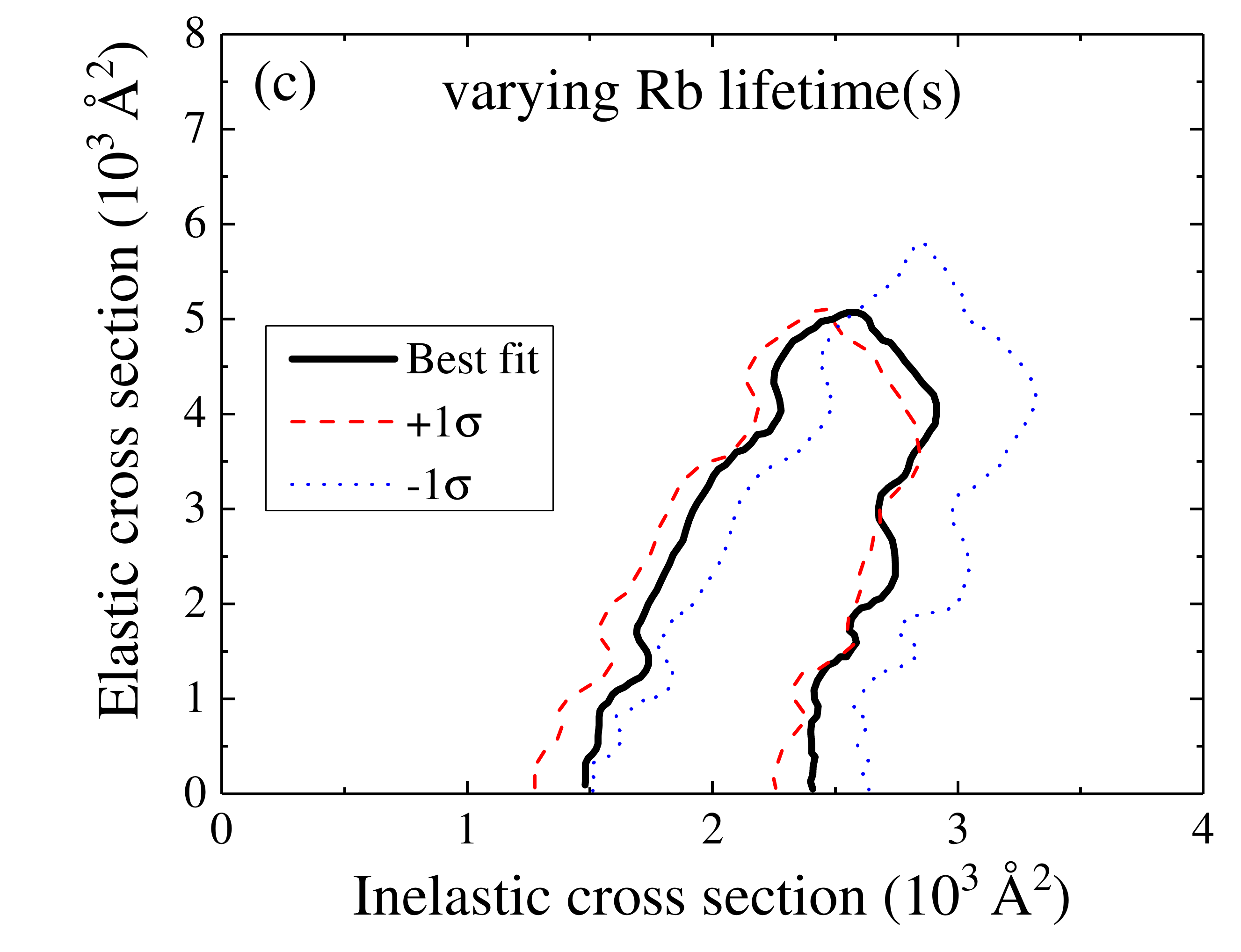}
\caption{Best-fit simulation results for variations in the Rb input parameters to the simulations.  All results are for the  $^{87}$Rb--$^{14}$ND$_{3}$ system.  The solid (black) lines correspond to using Rb parameters as determined by a best fit to the measured Rb cloud behavior.  The dashed red and blue curves correspond to altering those parameters for the collision simulations by +1$\sigma$ and -1$\sigma$, respectively.  (a) Results for varying initial Rb density, (b) Rb widths in all three spatial dimensions, and (c) Rb lifetimes.}
\label{Varying_Rb_Parameters}
\end{figure}

With such a heavy reliance on the use of simulations, it is crucial to test the robustness of the cross-section assignments to changes in simulation input parameters.  To explore this, we ran the collision simulations while varying input parameters of the trapped Rb cloud, including the time-dependent Rb density, temperature, and lifetime. We varied these values by $\pm$1$\sigma$, where $\sigma$ refers to the uncertainty or standard error in that parameter.  Results obtained by varying each of these parameters are shown in Fig.~\ref{Varying_Rb_Parameters}.  Note that the contour lines indicate a greater uncertainty when using input parameters that differ from the experimentally determined nominal values.  This is because, if the input parameters used are incorrect, the model cannot precisely reproduce the experimental dynamics. However, the central values do not shift appreciably, indicating the simulations are robust to small variations in the input parameters. 

\begin{figure}[!thb]
\centering
\includegraphics[width=\linewidth]{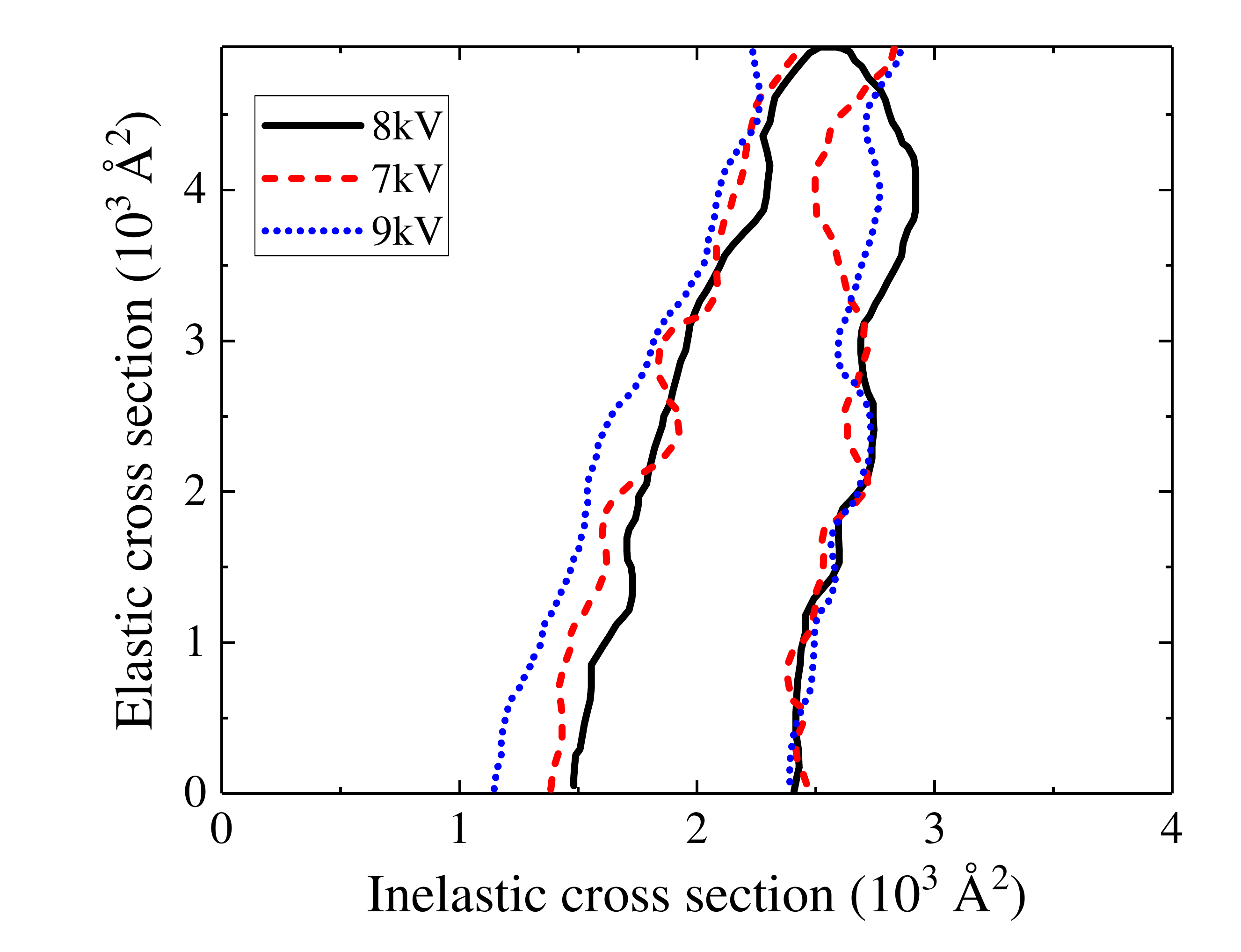}
\caption{Measured collision cross sections (1$\sigma$ confidence) for $^{87}$Rb-$^{14}$ND$_{3}$ while applying different voltages to the electrostatic trap electrodes.  Labels correspond to the magnitude of the voltage applied in the quadrupole configuration.}
\label{voltages}
\end{figure}

Because collisions between ND$_{3}$ and Rb are strongly affected by electric fields, it is also important to carry out the experiment with different average electric fields within the trap to verify our assumption that most molecules experience a strong enough electric field that the collision cross sections are essentially constant in the trap.  If this assumption is not valid, or the challenging theoretical calculations are not accurate, carrying out the experiment with different electrostatic trapping fields should display a systematic shift in the determined cross sections. However, since a significant trapping potential must be maintained in order to keep the ND$_3$ confined, the range over which we are able to explore this degree of freedom is limited.  To explore this, collisions between $^{87}$Rb and $^{14}$ND$_{3}$ are studied with 7, 8, and 9~kV amplitude electric potentials applied to the trap electrodes.  For each voltage, the time evolution of the Rb cloud is determined, and the entire set of both preliminary and collision simulations are run again.  Results of these measurements appear in Fig.~\ref{voltages}.  Here, the 1$\sigma$ curves are shown for the three different trapping potentials.  No significant shift of the measured inelastic cross section is observed, indicating that the electric field effect on the cross sections is either fully saturated or that we are unable to explore a large enough range of applied voltages to resolve any difference.    

\section{Conclusions}

In this work, we have demonstrated both the benefits and complications that come from using a dual-trap environment for collision studies.  The long interaction time is a crucial advantage for the experimental system explored here, as the densities and cross sections are too low to be feasibly investigated with beam-based approaches.  However, with the benefit of the long interaction time comes significant challenges, including effects of the applied electromagnetic fields on the collisions and the time evolution of the two trapped distributions.  These challenges can be overcome, as illustrated by techniques described in this paper.  In particular, the development of detailed and accurate simulations were instrumental in successfully assigning cross sections based on the observed behavior.  The resulting combination of experimental platform and post-analysis appears to be robust to uncertainties in the measured experimental parameters.  With confidence in the experimental and analytical approach, probing collisions using different isotopologues is potentially a powerful tool to reveal effects of an overall mass scaling in the particle interactions.  In this particular case, where interactions are averaged over the trap volume and thus occur at various energies and applied fields, no isotope effect was observed.    

\section{Acknowledgments}  
The authors would like to thank Jeremy Hutson and Piotr \.Zuchowski for theoretical calculations to understand the effects of electric fields on the collisions.  This work was supported by the National Science Foundation, PHY-1734006, Air Force Office of Scientific Research, and the Royal Society (URF\textbackslash R1\textbackslash 180578).

\bibliographystyle{apsrev4-1}
\bibliography{references}

\end{document}